\documentclass{jfm}
\usepackage{epsfig,psfig,epsf,graphics,color,rotate}

\title{A geostrophic-like model for large Hartmann number flows}
\author{Thierry Alboussi\`ere}
\affiliation{Laboratoire de G\'eophysique Interne et Tectonophysique, CNRS, Observatoire de Grenoble, Universit\'e Joseph Fourier \\ Maison des G\'eosciences, BP 53, 38041 Grenoble Cedex 9, France}
\date{\today}

\begin{document}
\bibliographystyle{alpha}

\maketitle

\begin{abstract}
A flow of electrically conducting fluid in the presence of a steady magnetic field has a tendency to become quasi two-dimensional, {\it i.e.} uniform in the direction of the magnetic field, except in thin so-called Hartmann boundary layers. The condition for this tendency is that of a strong magnetic field,  corresponding to large values of the dimensionless Hartmann number ($H\!a >> 1$). 
This is analogous to the case of low Ekman number rotating flows, with Ekman layers replacing Hartmann layers. This has been at the origin of the homogeneous model for flows in a rotating frame of reference, with its rich structure: geostrophic contours and shear layers of Stewartson [\cite{S57,S66}], Munk [\cite{munk}] and Stommel [\cite{stommel}].  
In magnetohydrodynamics, the characteristic surfaces introduced by Kulikovskii [\cite{kuli}] play a role similar to the role of the geostrophic contours. 
However, a general theory for quasi two-dimensional magnetohydrodynamics is lacking. In this paper, a model is proposed which provides a general framework for quasi two-dimensional magnetohydrodynamic flows. Not only can this model account for otherwise disconnected past results, but it is also used to predict a new type of shear layer, of typical thickness $H\!a^{-1/4}$. 
Two practical cases are then considered: the classical problem of a fringing 
transverse magnetic field across a 
circular pipe flow, treated by Holroyd and Walker [\cite{HetW}], and the problem of a rectangular cross-section duct flow in a slowly varying transverse magnetic field. 
For the first problem, the existence of thick shear layers of dimensionless thickness of order of magnitude  $H\!a^{-1/4}$ explains why the flow expected at large Hartmann number was not observed in experiments. The second problem exemplifies a situation where an analytical solution had been obtained in the past [\cite{WL72}] for the so-called ``M-shape'' velocity profile, which is here understood as an aspect of general quasi two-dimensional magnetohydrodynamics.

\end{abstract}

\section{Introduction}

The relative influence of Lorentz forces versus viscous forces is measured by the dimensionless Hartmann number $H\!a$. At large Hartmann number, the flow can be analyzed in terms of an inviscid core flow and viscous shear layers. This is very similar to the case of low Ekman number flows in a rotating frame of reference, for which two-dimensional models are commonly used.   
The tendency of flows of electrically conducting fluids under a DC magnetic field to become two-dimensional has been known since the work of Hartmann and Lazarus [\cite{h37,hl37}]. 
A number of analytical [\cite{shercliff,chetLu,Hunt}] and experimental [\cite{m53,b67}] studies have confirmed this trend in the laminar regime and even in the next stage of so-called two-dimensional turbulence [\cite{lehnert,kolesnikov}]. It was then natural to take advantage of the well-known structure of the Hartmann layers to build a model for the two-dimensional flow. This was done by Sommeria and Moreau [\cite{sm82}], in terms of the core velocity and pressure. There have been also other two-dimensional models using the electrical potential and pressure as main variables [\cite{Molo}]. However these models have all in common that the flow is supposed to be contained between two parallel planes with a uniform transverse magnetic field. This constitutes a very particular case as the subsequent developments show. 

The work of Kulikovskii [\cite{kuli}] brought more generality since he considered electrically insulating  cavities of arbitrary shapes in arbitrary non-uniform magnetic fields. He shows that there is still a kind of two-dimensionality as the variation of physical quantities like pressure or electric potential along the magnetic lines can be obtained from the equations of the core of the flow. In addition, he shows that the quantity $ \int ds/\| B\| $, integral on a magnetic line of the inverse of the magnetic field intensity, plays a crucial role. Surfaces made of magnetic lines with the same value for this integral are called characteristic surfaces: Kulikovskii shows that the electric current and velocity have a tendency to lie on these characteristic surfaces at large Hartmann number. 
These surfaces can be considered as surfaces of least resistance to the flow. A flow can be forced to cross them but this will cost more in terms of pressure gradient required than in the case of a flow along them\footnote{For the same velocity, the ratio of pressure gradient required to drive a flow across characteristic surfaces is of the order of a factor Hartmann times the pressure gradient necessary to drive a flow along these surfaces.}.   
In a second paper [\cite{kuli73}], Kulikovskii applies his asymptotic structure to particular configurations, showing the powerful predictions that can be made using the concept of characteristic surfaces.

Later, the ideas of Kulikovskii have been applied by Holroyd and Walker [\cite{HetW}] in the case of a fringing transverse magnetic field to a circular duct. The transverse magnetic field is uniform on both sides of a step change. The authors follow Kulikovskii and assume that in the non-uniform region, the flow follows strictly the characteristic surfaces and that on each side it would then come back towards a fully-developed regime on a much larger length scale $H\!a^{1/2}$, as shown from scaling analysis\footnote{This scaling appears for the first time in [\cite{WalkerLudford}].}. They then re-derive a two-dimensional model with uniform magnetic field but non-uniform depth of the cavity in the direction of the magnetic field, in terms of pressure and electric potential. This model is then solved using expansions in eigenfunctions on both sides, where pressure and potential are matched in the central non-uniform magnetic field region. In doing so, they obtain the large Hartmann number asymptotic limit of the duct flow with fringing magnetic field. Unfortunately, experiments performed under a large Hartmann number by Holroyd [\cite{holroyd79}], $H\!a = 523$ based on the radius of the pipe, showed that the observed flow was significantly different from the predicted one: it is modified in the same way as expected but much less spectacularly. The reason for this discrepancy could not be explained satisfactorily by invoking possible viscous effects in the core or inertial effects since the Hartmann  and interaction numbers were both large enough to discard such effects. 

In a second attempt, Hua and Walker [\cite{HuaWalker}] derive two-dimensional equations not only for variable depth but also for a non-uniform magnetic field. The assumption of so-called parallel magnetic lines is made, whereby the magnetic field is supposed to be pointing nearly in the same direction ($z$), although its intensity depends on the other two directions ($x$ and $y$). This model could be solved numerically without the need to match the pressure and potential in the non-uniform magnetic field region. The results were consistent with experimental observations at comparable values of the Hartmann number. Moreover, the numerical solution could be obtained for higher values of the Hartmann number up to $7000$. The authors remark that, even at this large value of Hartmann number, the computed flow is far from the asymptotic prediction of Holroyd and Walker. As they expect the asymptotic regime to be reached when the square root of the Hartmann number is large compared to unity they conclude rather unconvincingly 
that $\sqrt{7000} \simeq 84$ is not large enough compared to unity to approach the asymptotic regime. 

A closely related problem is that of a cylindrical duct with expansion in a uniform transverse magnetic field [\cite{WalkerLudford}]. The topology of the characteristic surfaces is identical to that of the duct of constant diameter with varying transverse magnetic field. This study is purely analytical, based on matching solutions on both sides of the region of varying diameter.  

Another important case where the characteristic surfaces play a crucial role is that of the so-called entrance problem for rectangular duct flows. When the flow enters a region of varying transverse magnetic field, the flow appears to concentrate along the sides of the duct parallel to the magnetic field, forming a so-called M-shape velocity profile. This was reported from experimental observations in a range of papers [\cite{sss70,bs66,holroyd79,holroyd2}]. 
For this configuration again, Walker and co-workers played a major role in the analysis of the flow [\cite{wlh72,WL72,sw99}]. 

Throughout this paper, the question of the relevance of a two-dimensional MHD 
model will be addressed.  It is hoped that the reader will be convinced that two-dimensional  
equations of the type of those presented in [\cite{HetW,HuaWalker}] do
 represent MHD flows correctly, but that their scaling analysis has not received
 enough attention. Such analysis provides a global understanding of MHD flows. In 
particular, it will be shown how some known results can be rederived ({\it e.g.} the
M-shape profile) and how an open question arising in the case of a fringing 
 transverse magnetic field to a circular duct can be clarified.
Also, the comparison with the case of rotating flows will be made. As the 
two-dimensional analysis is more advanced and better understood for rotating flows, this 
comparison will hopefully give support to the proposed MHD analysis. In addition, strong magnetic fields are becoming widely available for research and industrial purposes, and understanding the asymptotic limit of large Hartmann number flows is becoming increasingly useful in metallurgy or other processes involving liquid metals, like liquid metal cooling or spallation neutron sources. 

\section{Three-dimensional asymptotics}
\label{3d}

In this section the different types of three-dimensional boundary layers appearing in MHD flows or flows in rotating systems of reference are reviewed briefly. These well-known features are recalled here for future comparison in section \ref{2d} with typical two-dimensional structures and to stress the analogy between MHD and rotating flows: the theory of rotating flows is more advanced than that of MHD flows and the main point of this paper is to show that the diversity of scaling laws found in the homogeneous model of rotating flows also exists for MHD flows. 

\subsection{MHD three-dimensional asymptotics}
\label{3dasymp}

The dimensionless vector position ${\bf x}$, magnetic field ${\bf B}$, velocity field ${\bf u}$, electric current density ${\bf j}$, pressure field $p$ and electric potential field $\phi$ are derived from their corresponding dimensional quantities using the scales $H$, $B_0$, $\nu / H$, $\nu \sigma B_0 / H$, $\rho \nu ^2 / H^2$ and $\nu B_0$ respectively, where $H$ is a length-scale of the configuration, $B_0$ a typical value of the applied magnetic field, $\rho$ the density of the fluid, $\nu$ its kinematic viscosity and $\sigma$ its electrical conductivity. The magnetic field is imposed externally and the magnetic Reynolds number is assumed to be small so that the dimensionless magnetic field ${\bf B}$ is divergence-free and curl-free according to Maxwell's equations. Moreover, inertia is also neglected so that the dimensionless expressions of Navier-Stokes equation and Ohm's law take the following form: 
\begin{eqnarray}
{\bf 0} & = & - {\bf \nabla} p + H\! a^2 \, {\bf j} \times {\bf B} + {\bf \nabla}^2 {\bf u}, \label{ns} \\
{\bf j} & = & - {\bf \nabla} \phi + {\bf u} \times {\bf B}. \label{ohm}
\end{eqnarray}
In addition, the velocity field ${\bf u}$ and electric current density ${\bf j}$ must be divergence-free and on the boundary of the cavity enclosing the fluid, the velocity must be zero as well as the component of the electric current density parallel to the normal vector, since only electrically insulated cavities are considered in this paper. One could consider a field of volume forces in the Navier-Stokes equation, like buoyancy forces in the Boussinesq approximation, without affecting the analysis of the structure of the flow, provided this force field does not vary on a short length-scale. In the set of equations (\ref{ns}) and (\ref{ohm}), the Hartmann number $ H\! a = \sqrt{\sigma / \rho \nu} B_0 H$ is the single dimensionless parameter. 

Asymptotic analysis refers to the limit of infinite Hartmann number $ H\! a $. Since Hartmann, it has been known that the flow can be divided into so-called core regions of finite size and layers of small thickness. A simple way to put these structures in evidence is to perform a local analysis of the governing equations (\ref{ns}), (\ref{ohm}), so that one can consider that the magnetic field is uniform. Let us denote $z$ the local coordinate aligned with the magnetic field and $x$, $y$ other coordinates such that ($x$, $y$, $z$) is a direct orthonormal system of reference. Taking twice the curl of (\ref{ns}) and substituting ${\bf \nabla } \times {\bf j}$ using the curl of (\ref{ohm}) leads to the following equation for the velocity field:  
\begin{equation}
H\! a_l^2 \, \frac{ \partial ^2 {\bf u}}{\partial z^2} - \left( {\bf \nabla}^2 \right) ^2 {\bf u} = {\bf 0}, \label{eq-u}
\end{equation}
where $H\! a_l = H\! a \, \| {\bf B} \| $ represents the local Hartmann number. 
The electric current density ${\bf j}$ can be shown to obey the same equation. The three-dimensional structures of MHD flows can be derived from the differential operator\footnote{
For completeness, it is recorded that this operator can be decomposed into two operators as follows: 
$
H\! a_l^2 \, { \partial ^2 }/{\partial z^2} - \left( {\bf \nabla}^2 \right) ^2 = \left( H\! a_l \, { \partial }/{\partial z} - {\bf \nabla}^2 \right) \circ  \left( H\! a_l \, { \partial }/{\partial z} + {\bf \nabla}^2 \right),  
$ 
where each of these operators governs its corresponding Shercliff variable [\cite{shercliff}]. These variables are linear combinations of the velocity and (induced) magnetic field, although they are a different linear combination than Elsasser variables [\cite{elsasser}]. One must introduce another dimensionless number, the magnetic Prandtl number $Prm\, = \, \mu \sigma \nu $, where $\mu$ is the magnetic permeability. While Elsasser variables are ${\bf u} \pm {\bf A}$, with ${\bf A}$ the Alfv\'en velocity
${\bf B}/ \sqrt{\rho \mu }$, Shercliff variables are ${\bf u} \pm Prm^{-1/2} {\bf A}$. 
}
 $H\! a_l^2 \, { \partial ^2 }/{\partial z^2} - \left( {\bf \nabla}^2 \right) ^2$. 

\subsubsection{Core regions}
When the Hartmann number tends towards infinity, the bi-Laplacian term cannot compete in finite-size regions and one gets a region where the second derivative of the velocity vanishes. In the absence of a strong electric current along the magnetic lines, it can be shown that a two-dimensional velocity field develops in these regions of finite extent. Even in the general case of a non-uniform magnetic field, a kind of two-dimensional state exists, since the variations of velocity along the magnetic lines are readily obtained from the governing equations. 

\subsubsection{Hartmann layers}
Where the magnetic field is not parallel to the wall, a Hartmann boundary layer develops. Obviously, the direction of maximum variation is the normal direction to the wall ${\bf n}$ and equation (\ref{eq-u}) can be simplified into:
\begin{equation}
H\! a_l^2 \, ({\bf B}.{\bf n})^2 \, \frac{ \partial ^2 {\bf u}}{\partial n^2} -  \frac{ \partial ^4 {\bf u}}{\partial n^4} = {\bf 0}, \label{eq-hartmann}
\end{equation}
where $n$ is the distance to the wall. Hartmann layers are solutions to this equation and the tangent velocity varies exponentially in them on a typical length-scale $H\! a^{-1} \, ({\bf B}.{\bf n})^{-1}$. 

\subsubsection{Parallel layers}
Viscous terms represented by the bi-Laplacian term in ({\ref{eq-u}) can also play a role in thin regions (boundary or free-shear layers) containing the magnetic field direction. In this case, equation (\ref{eq-u}) takes the following form: 
\begin{equation}
H\! a_l^2 \, \frac{ \partial ^2 {\bf u}}{\partial z^2} -  \frac{ \partial ^4 {\bf u}}{\partial n^4} = {\bf 0}, \label{eq-parallel}
\end{equation}
where $n$ is again a coordinate perpendicular to the layer. The solutions to this equation are parallel layers. It is difficult to obtain a general solution in these layers, but a scaling analysis shows that if these layers stretch along a distance of order unity along the magnetic field lines, their thickness must be of order $H\! a_l^{-1/2}$. Parallel layers were first analyzed by Shercliff [\cite{shercliff}]. 

Interestingly, another result can be derived from equation (\ref{eq-parallel}) in the case when the cavity is of infinite length in the direction of the magnetic field. This gives a limit to the size of the core region where the flow is two-dimensional. There can be variations along the magnetic lines on a large length-scale: if the typical cavity length scale perpendicular to the magnetic lines is of order $1$, both terms in (\ref{eq-parallel}) can balance provided the length-scale along the magnetic lines is of order $H\! a_l$. This situation arises typically when an object of size of order unity is placed in an infinite region: its effect on a flow around it stretches along the magnetic lines as far as a distance of order $H\!a$.

In the case of parallel layers and when the magnetic field is non uniform, the local analysis is not rigorous as one has to consider variations of the velocity on a length-scale of order unity or more along magnetic lines. Equation (\ref{eq-parallel}) is therefore not correct. However the scaling analysis remains valid: in a paper by Todd [\cite{todd68}], the rigorous analysis of parallel layers developing along curved magnetic lines is performed and it is shown that these layers are still lying on magnetic lines and are still of typical thickness $H\!a^{-1/2}$. 

\subsubsection{Roberts layers}
Roberts layers develop as a curved wall approaches a point where it becomes tangent to the magnetic lines. The Hartmann layer solution presents a singularity as its thickness diverges. The divergence is resolved by a so-called Roberts layer [\cite{roberts67}]. If the curvature is of order unity at the point of contact, a small region elongated in the magnetic field direction must satisfy the geometric constraint that its thickness across magnetic lines scales as the square of its length along these lines. By scaling analysis of equation (\ref{eq-parallel}) the only possibility is a thickness of order $H\! a_l^{-2/3}$ and a length of order $H\! a_l^{-1/3}$ along the magnetic lines. 

\begin{figure}
\label{3dstruct}
\begin{center}
\input{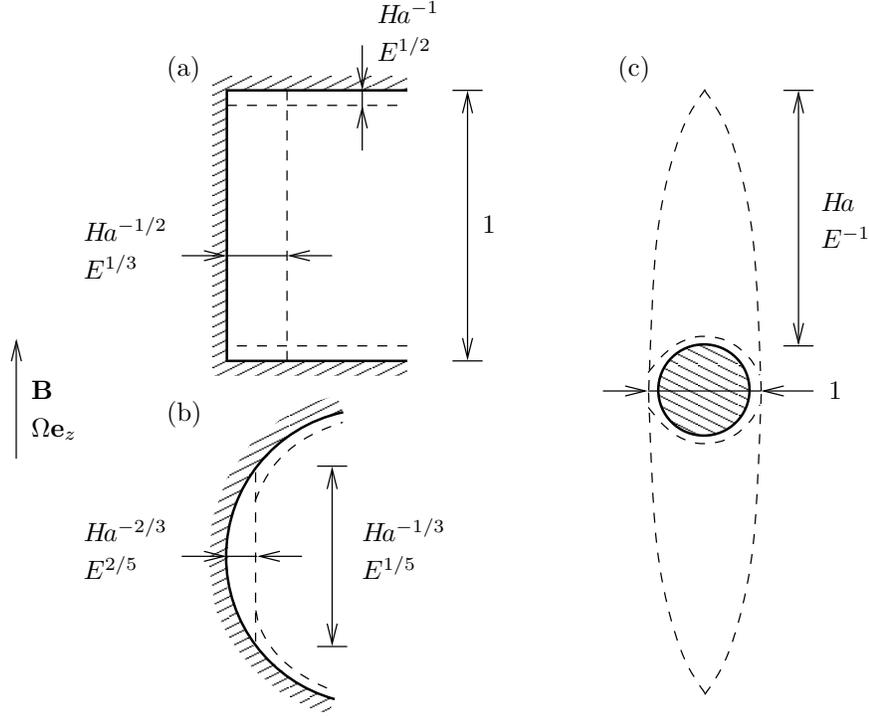}
\caption{Sketch of three-dimensional structures in MHD and rotating flows: (a) Hartmann~/~Ekman layers and MHD parallel layer~/~inner Stewartson layer; (b) Roberts layer~/~singular Ekman layer; (c) MHD wake~/~`Taylor' column.}
\end{center}
\end{figure}

\subsection{Rotating flows three-dimensional asymptotics}

Now, instead of applying a steady magnetic field ${\bf B}$, it is supposed that the reference system considered is rotating with the rate $\Omega$ along the $z$ axis. A Coriolis force replaces the Lorentz force in the momentum equation. 
The same dimensionless variables are used for distance, velocity and pressure as in section \ref{3dasymp}. The governing equation can then be written as follows: 
\begin{eqnarray}
{\bf 0} & = & - {\bf \nabla} p + 2 \, E^{-1} \, {\bf u} \times {\bf e}_z + {\bf \nabla}^2 {\bf u}, \label{nsrota} 
\end{eqnarray}
where $E=\nu / (\Omega H^2)$ is the dimensionless Ekman number. Taking the curl of equation (\ref{nsrota}) leads to an equation involving ${\bf u}$ alone:
\begin{eqnarray}
{\bf 0} & = 2 \, E^{-1} \, \frac{\partial {\bf u}}{\partial z} + {\bf \nabla}^2  ({\bf \nabla} \times {\bf u}). \label{curlnsrota} 
\end{eqnarray}
One can even differentiate this equation with respect to $z$ to eliminate the curl operator and obtain a form suitable for analytical investigations:
\begin{eqnarray}
{\bf 0} & = 4 \, E^{-2} \, \frac{\partial ^2 {\bf u}}{\partial z^2} +  ({\bf \nabla}^2 )^3 {\bf u}. \label{dzcurlnsrota} 
\end{eqnarray}
Under the form (\ref{curlnsrota}) or (\ref{dzcurlnsrota}), one can extract the possible structures for the velocity field using scaling analysis (see [\cite{greenspan}] for a classical treatment). When a boundary layer develops on a wall that does not contain the direction of rotation, a layer of dimensionless thickness $E^{1/2}$ develops: this is an Ekman layer and corresponds to the Hartmann layer in MHD. It can also be seen that shear layers containing the direction of rotation can develop. If the dimensionless length of the cavity in the direction of the rotation is of order one, these layers must have a thickness of order $E^{1/3}$: they are the strict equivalent to the MHD parallel layers. When an object of dimensionless size one is held in place in an infinite domain, then its `wake' extends a distance $E^{-1}$ in the direction of the rotation. Finally, similar regions to Roberts layers develop at the point of a curved wall where it becomes tangent to the direction of rotation: these regions have a typical length-scale of order  $E^{1/5}$ and $E^{2/5}$ in the direction and perpendicular to the direction of rotation respectively, if the curvature is of order unity. They correspond to a singularity of diverging Ekman layers.
 
At this point, it can be seen that there is a close correspondence between the possible structures in MHD and in rotating flows according to the three-dimensional equations (see Fig.~1 
for a pictorial summary). However, it is not clear how one can possibly extract the typical $E^{1/4}$ main component of Stewartson's layers from the scaling of the differential operator $4 \, E^{-2} \, {\partial ^2 }/{\partial z^2} +  ({\bf 
\nabla}^2 )^3 $ appearing in equation (\ref{dzcurlnsrota}). This becomes obvious when two-dimensional equations are derived.  

\section{Two-dimensional models}
\label{2d}

A cavity of dimensionless length-scale unity in the direction of the imposed magnetic field or rotation is considered. As we have seen above in section \ref{3d}, the largest three-dimensional structures in the direction perpendicular to the magnetic field or rotation direction are of order $H\!a^{-1/2}$ or $E^{1/3}$. Hence, if only larger perpendicular scales are considered, the flow consists simply of a two-dimensional core bounded by Hartmann or Ekman layers. This is at the origin of the two-dimensional models.

\subsection{MHD two-dimensional model}
\label{2dmag}

In most cases, MHD core flows are two-dimensional, except when the forcing and boundary conditions are such that an  odd velocity profile is generated along a magnetic field line (a linear variation in the core). In this case, the MHD braking is very strong: this is the case referred to as `singular symmetry' in [\cite{PhysA}]. The other symmetry called `regular symmetry' corresponds to an even velocity profile (which is nearly uniform in the core) along magnetic lines, and the MHD braking is $H\!a$ times weaker. Hence, the general case bears resemblance with the `regular symmetry' and corresponds indeed to the idea of a two-dimensional flow. 

In the following analysis a cavity of general shape and a non-uniform magnetic field are considered. 
Following most authors, the approximation of straight magnetic lines is made here. In this approximation, the curvature of the magnetic lines is ignored, the intensity of the magnetic field is taken as constant on each magnetic line but can vary from one line to another. This is clearly an approximation intended to simplify the analytical calculations, but it has been shown to be safe in the few studies where this approximation was not made [\cite{kuli,todd68,PhysA}]. This approximation does not affect the qualitative features of the flows, although the imposed magnetic fields are not physical since they are rotational. 

As shown in Fig.~2,  
the geometry of the electrically insulated cavity consists of the space situated between two surfaces, an upper surface ${\cal{S}}^u$ and a lower surface ${\cal{S}}^l$, defined in a orthonormal coordinate system $(x,y,z)$ by the two functions $z^u (x,y)$ and $z^l (x,y)$ respectively. The $z$ axis is chosen to coincide with the magnetic field direction, in the straight magnetic lines approximation. Its intensity depends on $x$ and $y$, ${\bf B}=B_z(x,y){\bf e}_z $. The functions $z^u$, $z^l$ and $B_z$ are dimensionless functions of the dimensionless coordinates $x$ and $y$, where the arbitrary scales $H$ and $B_0$ for length and magnetic field intensity defined in section \ref{3d} continue to be used. 

\begin{figure}
\begin{center}
\input{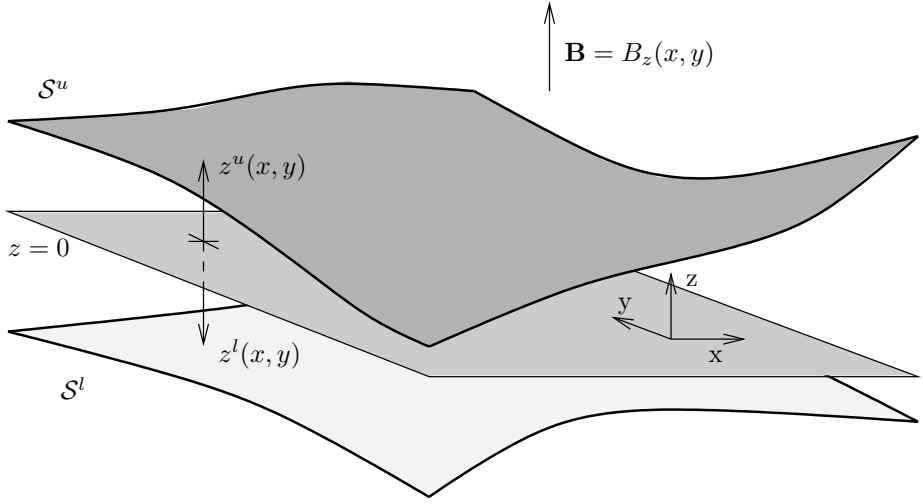}
\caption{Geometry and magnetic field for two-dimensional analysis}
\end{center}
\label{schema}
\end{figure}

It is assumed that, in the core of the flow, the velocity and electric current density components perpendicular to the magnetic field  are independent of $z$. Using for instance the value of these fields on a reference plane $z=0$, one can define the two-dimensional fields: 
\begin{equation}
{\bf u}_0 = \left[ \begin{array}{l} u_{0x} (x,y) \\ u_{0y} (x,y) \end{array} \right], \ \ \ \  
{\bf j}_0 = \left[ \begin{array}{l} j_{0x} (x,y) \\ j_{0y} (x,y) \end{array} \right]. \label{u0j0}
\end{equation}
Pressure and electrical potential are also independent of the magnetic field direction $z$ in the core, and their two-dimensional representatives are denoted $p_0$ and $\phi _0$. Considering the restricted versions of the general three-dimensional equations (\ref{ns}) and (\ref{ohm}) in the plane ($x,y$) in the core, two-dimensional equations governing ${\bf u}_0$, ${\bf j}_0$, $p_0$ 
and $\phi _0$ are obtained:
\begin{eqnarray}
{\bf 0} & = & - {\bf \nabla} p_0 + H\! a^2 \, B_z \, {\bf j}_0 \times {\bf e}_z + {\bf \nabla}^2 {\bf u}_0 , \label{ns0} \\
{\bf j}_0 & = & - {\bf \nabla} \phi _0 + B_z \, {\bf u}_0 \times {\bf e}_z. \label{ohm0}
\end{eqnarray}
These governing equations are not sufficient to describe the MHD two-dimensional flows. First, one needs to take into account the global conservation of mass and of electric charge. Second, one needs to take into account the contribution due to the Hartmann layers in these conservation relationships. 

The mass and charge conservation laws will be expressed in terms of global two-dimensional mass and electric charge flux densities ${\bf Q}$ and ${\bf I}$, taking into account the core and Hartmann layers contributions. The core contribution is obtained by multiplying the two-dimensional vector fields ${\bf u}_0$ and ${\bf j}_0$ by the length of the magnetic line within the cavity $z^u - z^l$. The Hartmann contribution to the mass flux is neglected as this corresponds simply to a deficit, as measured by the displacement thickness of the Hartmann boundary layer. It introduces an error of order $H\!a^{-1}$, comparable to the approximations already made when assuming a two-dimensional core flow. On the contrary, the Hartmann layer contribution to the electric charge flux is fundamental in MHD. It is proportional to the core velocity adjacent to the Hartmann layer and in the direction perpendicular to both this core velocity and the wall normal vector. The core velocity adjacent to the Hartmann layer ${\bf u}_c$ must be tangent to the wall and the integral of the electric current over the Hartmann layer thickness ${\bf I}_{H\!a}$ depends on the wall normal unit vector ${\bf n}$ and ${\bf u}_c$ (see [\cite{HuetSh}]): 
\begin{equation}
{\bf I}_{H\!a} = - sign({\bf B}.{\bf n}) \, H\!a^{-1} \, {\bf u}_c \times {\bf n}. \label{Iha}
\end{equation}
This surface electric current density needs to be expressed in terms of the coordinates $x$ and $y$ which will be used in the following two-dimensional analysis. To this end, let us calculate the amount of electric current flowing across a small line element spanned by $dx $ and $dy$ (see Fig.~3).
This line element corresponds to a line element $d{\bf l} $ on the tangent plane to the surface 
 and, from the components $u_{0x}$ and $u_{oy}$ of ${\bf u}_0$, one can also build the tangent core velocity:
\begin{equation}
d {\bf l} = \left[ \begin{array}{l} dx \\ dy \\ dx \frac{\partial z^u}{\partial x} + dy \frac{\partial z^u}{\partial y}
\end{array} \right], \ \ \ \
{\bf u}_c = \left[ \begin{array}{l} u_{0x} \\ u_{0y} \\ u_{0x} \frac{\partial z^u}{\partial x} + u_{0y} \frac{\partial z^u}{\partial y}
\end{array} \right], \label{dluc}
\end{equation}
where the upper surface ${\cal{S}}^u$ is considered here, the treatment for the lower surface being similar. 
The amount of electric current flowing across the line element $d{\bf l}$ is proportional to the magnitude of the vector $d{\bf l} \times ( {\bf u}_c \times {\bf n} )$. This vector can be shown to be equal to $( d {\bf l} . {\bf u}_c ) \, {\bf n}$. Hence, with a factor $H\!a^{-1}$, the flux of electric current through $d{\bf l}$ is given by:
\begin{equation}
dx \left[ u_{0x} + u_{0x} \left( \frac{\partial z^u}{\partial x} \right) ^2 + u_{0y} \frac{\partial z^u}{\partial x} \frac{\partial z^u}{\partial y} \right] + 
dy \left[ u_{0y} + u_{0y} \left( \frac{\partial z^u}{\partial y} 
\right) ^2 + u_{0x} \frac{\partial z^u}{\partial x} \frac{\partial z^u}{\partial y}
\right]. \label{flux-line}
\end{equation}
The two-dimensional vector flux ${\bf I}_{H\!a}^u$ in the $(x,y)$ coordinate system that represents that electric current flux has components: 
\begin{equation}
{\bf I}_{H\!a}^u = \left[ \begin{array}{l} I_{H\!ax}^u \\ I_{H\!ay}^u \end{array} \right] = H\!a^{-1} \left[ \begin{array}{l}  - u_{0y} - u_{0y} \left( \frac{\partial z^u}{\partial y}            
\right) ^2 - u_{0x} \frac{\partial z^u}{\partial x} \frac{\partial z^u}{\partial y}
 \\ u_{0x} + u_{0x} \left( \frac{\partial z^u}{\partial x} \right) ^2 + u_{0y} \frac{\partial z^u}{\partial x} \frac{\partial z^u}{\partial y}   \end{array} \right].
\end{equation}
The lower surface produces a similar electric current flux. Both contributions are expressed by a two-dimensional tensor, ${\cal{F}}^u$ and ${\cal{F}}^l$ for the upper and lower surfaces respectively:
\begin{equation}
{\bf I}_{H\!a}^u = H\!a^{-1} {\cal{F}}^u . {\bf u}_0, \ \ \ \ 
{\bf I}_{H\!a}^l = H\!a^{-1} {\cal{F}}^l . {\bf u}_0, \label{Iha-ul}
\end{equation}
with 
\begin{equation}
{\cal{F}}^u = \left[ \begin{array}{ll} - \frac{\partial z^u}{\partial x} \frac{\partial z^u}{\partial y} & -1 -\left( \frac{\partial z^u}{\partial y}
\right) ^2 \\ 
1 + \left( \frac{\partial z^u}{\partial x} \right) ^2 & \frac{\partial z^u}{\partial x} \frac{\partial z^u}{\partial y} \end{array} \right], \ \ \ \ 
{\cal{F}}^l = \left[ \begin{array}{ll} - \frac{\partial z^l}{\partial x} \frac{\partial z^l}{\partial y} & -1 -\left( \frac{\partial z^l}{\partial y}
\right) ^2 \\ 
1 + \left( \frac{\partial z^l}{\partial x} \right) ^2 & \frac{\partial z^l}{\partial x} \frac{\partial z^l}{\partial y} \end{array} \right]. \label{ftensors}
\end{equation}

\begin{figure}
\begin{center}
\input{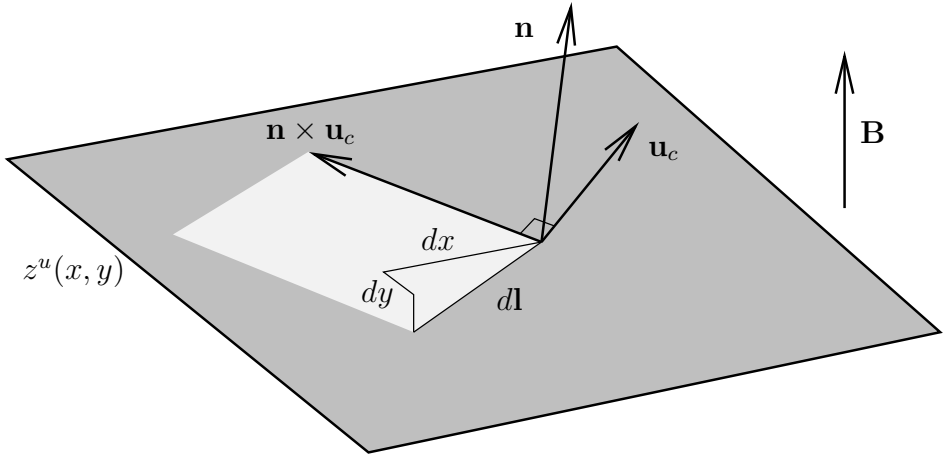}
\caption{Local tangent wall surface and relationship between the core velocity and total electric current flowing in the Hartmann layer}
\end{center}
\label{geomflux}
\end{figure}

It is now possible to express the total mass and charge two-dimensional fluxes, including the core and Hartmann layers contributions: 
\begin{equation}
{\bf Q} = (z^u - z^l) {\bf u}_0, \ \ \ \ 
{\bf I} = (z^u - z^l) {\bf j}_0 + {\bf I}_{H\!a}^u + {\bf I}_{H\!a}^l. \label{fluxes} 
\end{equation}
For impermeable and electrically insulated surfaces, these two-dimensional flux densities are divergenceless, hence can be expressed in terms of the streamfunctions $\psi$ and $h$. 
\begin{equation}
{\bf Q} = {\bf \nabla} \psi \times {\bf e}_z,  \ \ \ \ 
{\bf I} = {\bf \nabla} h \times {\bf e}_z. \label{psi-h}
\end{equation}
Taking the curl of the two-dimensional equations (\ref{ns0}) and (\ref{ohm0}) eliminates pressure and electric potential fields and yields the following equations, which have only one component along $z$:
\begin{eqnarray}
{\bf 0} & = & - H\!a^2\, \left[ {\nabla }.(B_z {\bf j}_0 ) \right] {\bf e}_z + {\bf \nabla} ^2 \left( {\bf \nabla } \times {\bf u}_0 \right) , \label{curlns} \\
{\bf \nabla } \times {\bf j}_0 &=& - \left[ {\nabla }. (B_z {\bf u}_0 ) \right] {\bf e}_z. \label{curlohm}
\end{eqnarray}
These equations can then be expressed entirely in terms of $\psi $ and $h$, using (\ref{psi-h}), (\ref{fluxes}) and (\ref{Iha-ul}) for substitution. For the sake of concise expressions, let us introduce $d = z^u - z^l$ the depth of the cavity in the direction of the magnetic field and ${\cal{F}} = {\cal{F}}^u + {\cal{F}}^l$ the sum of the upper and lower Hartmann electric currents in terms of the core two-dimensional velocity. During the substitution, the contribution of these Hartmann currents to equation (\ref{curlohm}) is neglected, while their contribution to (\ref{curlns}) is retained\footnote{The motivation for these different treatments is that the Hartmann currents are nearly curl-free while their divergence can be significant.}:
\begin{eqnarray}
H\!a \, {\bf \nabla}.\left( \frac{B_z}{d^2} {\cal{F}}. ({\bf \nabla} \psi  \times {\bf e}_z) \right) {\bf e}_z & = & H\!a^2 \, {\bf \nabla} \left( \frac{B_z}{d} \right) \times {\bf \nabla} h  + {\bf \nabla} ^2 \left[ {\bf \nabla} . \left( \frac{{\bf \nabla } \psi }{d} \right) \right] {\bf e}_z, \label{eqpsi}\\
{\bf \nabla}. \left( \frac{{\bf \nabla } h }{d} \right) {\bf e}_z 
&=& {\bf \nabla} \left( \frac{B_z}{d} \right) \times {\bf \nabla} \psi. \label{eqh}
\end{eqnarray}
Both equations have almost a symmetrical structure. The curl of Navier-Stokes (\ref{eqpsi}) has a diffusion term for $\psi$ on the left-hand side: it is related to the two-dimensional vorticity of the flow. The equation has a term of the form ${\bf \nabla} K \times {\bf \nabla} h $ on the right-hand side, with $K=B_z / d$. This fundamental term corresponds to the flow of total electric current crossing the characteristic surfaces. Finally, the last bi-Laplacian term on $\psi$ represents the effect of viscosity in the two-dimensional flow. This last term will be shown to be negligible in most cases, and then 
equation (\ref{eqpsi}) expresses that when the electrical current goes across characteristic surfaces, this creates vorticity. 
Equation (\ref{eqh}) is similar, except that there is no equivalent to the last term of (\ref{eqpsi}). It says that when the flow goes across characteristic surfaces, this creates a region of curl for the electric current. 

It is important to stress that the characteristic function $K = B_z / d$ appears naturally in these two-dimensional equations, as could be expected from previous analyses [\cite{kuli,PhysA}]. The derivation and results presented above are close to those presented in [\cite{a2001}]. This analysis is also very close to the analysis by Holroyd and Walker [\cite{HetW}] and Hua and Walker [\cite{HuaWalker}], where the main variables are the pressure and electric potential instead of the streamfunctions $\psi$ and $h$ considered here. Also, they have neglected from the beginning the viscous friction in the two-dimensional flow. Although the analysis above about the significance of the terms is not carried out in papers [\cite{HetW,HuaWalker}], it can be checked after rearranging different terms that equations (17) and (18) in [\cite{HuaWalker}] display the same structure as (\ref{eqpsi}) and (\ref{eqh}) of this paper, with the characteristic function $B_z / d$ playing an important role. 

Equations (\ref{eqpsi}) and (\ref{eqh}) are going to be solved rigorously in sections \ref{fringing} and \ref{entry}. Nevertheless, it is important to perform a scaling analysis of these equations to identify the possible structures that can be expected in two-dimensional flows. For the purpose of the scaling analysis, the depth $d$ and magnetic field $B_z$ can be considered to be uniform and unity in all terms except for the two terms involving the gradient of $K$, since variation of $K=B_z / d$ is crucial. Moreover, in the spirit of a local analysis, a change of coordinates is introduced, in which $(x,y)$ is replaced by orthonormal intrinsic coordinates $(r,s)$, where $r$ varies in the direction perpendicular to the characteristic surfaces, while $s$ varies along the characteristic surfaces. The symbol $G$ is chosen to denote the magnitude of the gradient of $K$: $G = \| {\bf \nabla} (B_z / d) \| $. With these simplifying assumptions and notations, equations (\ref{eqpsi}) and (\ref{eqh}) take the form:
\begin{eqnarray}
H\!a \, {\nabla}^2 \psi & = & H\!a^2 \, G \frac{\partial h}{\partial s}  + ( {\nabla} ^2 ) ^2 \psi , \label{eqpsi2}\\
{\nabla} ^2  h 
&=& G \frac{\partial \psi }{\partial s}  \label{eqh2}
\end{eqnarray}
Taking the Laplacian of (\ref{eqpsi2}) and substituting $h$ using (\ref{eqh2}) provides the fundamental differential equation governing the two-dimensional streamfunction:
\begin{equation}
H\!a \,  ({\nabla}^2 )^2 \psi = H\!a^2 \, G^2 \, \frac{\partial ^2 \psi}{\partial s^2} + ({\nabla}^2 )^3 \psi. \label{eqpsi3}
\end{equation}
Under this form, it is convenient to perform a scaling analysis of the two-dimensional MHD flows. First of all, in the absence of thin layers, the term with the largest power of $H\!a$ is dominant. Consequently the second variations of $\psi$ along the characteristic surfaces must be zero. In fact, very often the first variation of $\psi$ will be zero along theses surfaces, in order to minimize Ohmic dissipation (see equation (\ref{eqh2})). This corresponds to the `core regions' of two-dimensional flows. 
Different type of layers can appear, depending on whether they develop in a direction parallel to characteristic surfaces or not. If they develop along a wall not parallel to the characteristic surfaces, then equation (\ref{eqpsi3}) becomes:
\begin{equation}
H\!a \,  \frac{\partial ^4 \psi}{\partial n^4} = H\!a^2 \, G^2 \, {\rm cos}^2 \theta \,  \frac{\partial ^2 \psi}{\partial n^2} + \frac{\partial ^6 \psi}{\partial n^6}, \label{eqpsi4}
\end{equation}
where $n$ is the normal direction to the wall and $\theta$ is the angle between the normal to the wall and the direction of characteristic surfaces. The thinnest structure that can emerge from equation (\ref{eqpsi4}) is obtained when the last term (viscous core friction) is balanced with the first term (Hartmann layer friction). This leads to a typical thickness of $H\!a^{-1/2}$ and must immediately be discarded as this scale corresponds to the scale of the three-dimensional parallel layers.
Indeed, the three-dimensional analysis in section \ref{3dasymp} shows that for a length scale of order $H\!a^{-1/2}$ or less in the direction perpendicular to the magnetic field, the variations of velocity in the core of the flow are of order unity in the direction of the magnetic field: hence the two-dimensional model ceases to be valid. 
This means that the core friction will always be negligible compared to Hartmann layer friction at larger scales. The other possibility is to balance Hartmann friction with the topographic term: this leads to a layer of typical thickness $H\!a^{-1/2} G^{-1} {\rm cos}^{-1} \theta$. Again, if $G$ and ${\rm cos} \theta$ are of order unity, this two-dimensional layer is not physical as it is replaced by a $H\!a^{-1/2}$ three-dimensional parallel layer. However, if $G \, {\rm cos} \theta$ is very small compared to unity, this thickness $H\!a^{-1/2} G^{-1} {\rm cos}^{-1} \theta$ is larger than the parallel layers and constitutes a valid two-dimensional layer. This type of layer has been analyzed by Walker and Ludford [\cite{WalkerLudford}] who started from the three-dimensional equations. These layers will be examined in section \ref{entry} from the easier point of view of the two-dimensional equations. 
They are similar to Stommel layers [\cite{stommel}] in rotating flows, as the main balance involves topographic effects and friction in Ekman layers. However, as will be seen in the next section, the Stommel solution does not apply, as the core viscous friction turns out to be larger than the Ekman friction in rotating flows. Hence, Stommel layers are replaced by thicker Munk layers [\cite{munk}]. 

If one considers now a layer developing along a characteristic surface, equation (\ref{eqpsi3}) can be written: 
\begin{equation}
H\!a \,  \frac{\partial ^4 \psi}{\partial r^4} = H\!a^2 \, G^2 \, \frac{\partial ^2 \psi}{\partial s^2} + \frac{\partial ^6 \psi}{\partial r^6}, \label{eqpsi5}
\end{equation}
Again, it can be seen that the last core friction term will always be negligible on scales larger than $H\!a^{-1/2}$ and can thus be ignored in this two-dimensional analysis. 

The scaling analysis depends on the scale along the characteristic surface. If this scale is of order unity ($\partial / \partial s \sim 1$), then the balance between the Hartmann friction term and the topographic term leads to a layer of typical thickness $H\!a^{-1/4} G^{-1/2}$ which is much thicker than parallel layers even if $G$ is of order unity. This type of layer will be discussed in section \ref{fringing}. 

Finally, if the cavity is infinite in the direction of the characteristic surfaces and of size unity in the perpendicular direction, equation (\ref{eqpsi5}) can provide the typical long length-scale of development of structures along characteristic surfaces. If $\partial / \partial r \sim 1$, then the balance between Hartmann friction and topographic effects leads to a scale $H\!a^{1/2} G$. Walker and co-workers [\cite{WalkerLudford}] had found that a distance of order $H\!a^{1/2}$ was the length-scale necessary from the position where the magnetic field was varying to reach a fully-developed flow in a pipe with transverse magnetic field. 

There can be other possibilities like the case when a wall would become tangent to the characteristic surfaces: similarly to Roberts layers, they would correspond to a singularity in the $H\!a^{-1/2} G^{-1}$ layer. For a curvature of order unity, scaling analysis of (\ref{eqpsi5}) when the length-scale perpendicular to the wall scales as the square of the length-scale along the characteristic surfaces, one gets a region of size $H\!a^{-1/3}G^{-2/3}$ by $H\!a^{-1/6}G^{-1/3}$.
 The structures discussed above are believed to be the most typical ones and are illustrated in Fig.~4.

As discussed above, the core viscous term $\left( \nabla ^2 \right) ^2 \psi $ in equation (\ref{eqpsi2}) is always negligible when the two-dimensional equations are valid. Without this term, equation (\ref{eqpsi2}) may be multiplied by $H\!a^{-3/2}$, then successively added to and subtracted from equation (\ref{eqh2}) providing the following equations:
\begin{equation}
{\nabla} ^2 \left(\psi / H\!a^{1/2} \pm  h \right)
= \pm H\!a^{1/2} G \frac{\partial }{\partial s} \left( \psi / H\!a^{1/2} \pm h  \right).  \label{pseudo-shercliff}
\end{equation}
Those familiar with Shercliff variables will notice that $\psi / H\!a^{1/2} \pm h$ play here a corresponding role. The privileged direction is that of characteristic surfaces instead of the direction of the magnetic field for Shercliff variables. This may help to accept the typical length-scales derived above. This also indicates that the magnitude of the electric current ($h$) is of order $H\!a^{-1/2}$ that of the velocity ($\psi$) in general. 

\begin{figure}
\begin{center}
\input{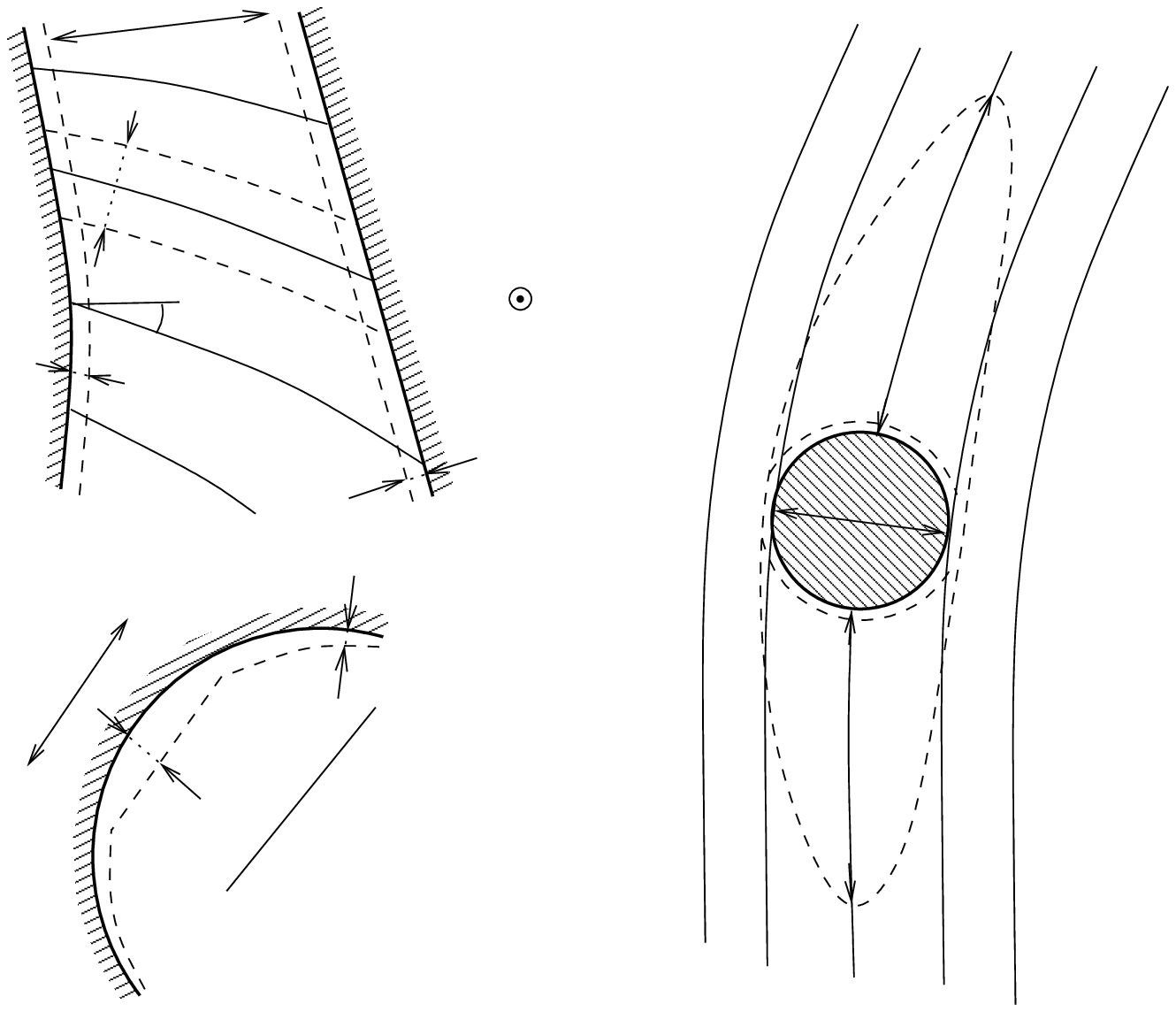}
\caption{Two-dimensional MHD structures.}
\end{center}
\label{twodmhd}
\end{figure}

\subsection{The homogeneous model of rotating flows}

The corresponding two-dimensional model in the case of rotation is called the the homogeneous model. Its derivation can be found in many textbooks ({\it{e.g.}} Greenspan [\cite{greenspan}]) and is sketched here in order to emphasize the similarity with the MHD model. At the upper and lower surfaces, the role of the Ekman layer is now to produce a cross-flow with respect to the core flow direction. The integral of this cross flow scales as $E^{1/2}$ times the core velocity. For this term and for the purpose of this approximate derivation, the angles between the direction of rotation and the wall normals are neglected. The total vertically-integrated volume flow rate can then be expressed:
\begin{equation}
{\bf Q} = d \, {\bf u}_0 + E^{1/2} {\bf e}_z \times {\bf u}_0. \label{mass-rot}
\end{equation}
The core two-dimensional equation is now obtained by restricting equation (\ref{nsrota}) to the $(x,y)$ components in the core of the flow:
\begin{equation}
{\bf 0} = -{\bf \nabla } p_0 + 2\, E^{-1} {\bf u}_0 \times {\bf e}_z + {\bf \nabla} ^2 {\bf u}_0 . \label{nsrota2d}
\end{equation}
Its curl takes the form: 
\begin{equation}
{\bf 0} = - 2\, E^{-1} ({\bf \nabla} . {\bf u}_0 ) {\bf e}_z + {\bf \nabla} ^2 ( {\bf \nabla } \times {\bf u}_0 ) . \label{curlrota2d}
\end{equation}
Again, one can use the streamfunction $\psi$, defined in (\ref{psi-h}) and substitute it for ${\bf u}_0$ in (\ref{curlrota2d}). The Ekman layer cross-flow is neglected in the viscous term and plays a crucial role in the other term. This results in the homogeneous model: 
\begin{equation}
2 \, E^{-1/2} \, {\bf \nabla} . \left( \frac{{\bf \nabla} \psi }{d} \right) {\bf e}_z = 2\, E^{-1} \, {\bf \nabla} \left( \frac{1}{d} \right)  \times {\bf \nabla} \psi + {\bf \nabla }^2 \left( {\bf \nabla} . \left( \frac{{\bf \nabla} \psi }{d} \right) \right) \, 
{\bf e}_z. \label{homogeneous}
\end{equation}

A similar change of two-dimensional coordinates to that in the previous section is operated, where the new $(r,s)$ system is defined such that $r$ is constant along lines of constant depth, the geostrophic contours, and $s$ is changing along these lines. Defining $\beta = \| {\bf \nabla} ( 1 / d ) \| $ in accordance to the so-called $\beta$-plane approximation{\footnote{The traditional $\beta$-plane approximation corresponds to the linearized variation of latitude, but this is equivalent to a change in depth, as far as the two-dimensional model is concerned.}}, equation ({\ref{homogeneous}) can be written locally: 
\begin{equation}
2 \, E^{-1/2} \, {\bf \nabla} ^2 \psi = 2\, E^{-1} \, \beta \frac{\partial  \psi}{\partial s} + ({\bf \nabla }^2)^2  \psi , \label{eq-psi-rot}
 \label{homogeneous2}
\end{equation}
under the assumption that $d$ is of order unity. Although the order of derivation of the terms and the powers of Ekman number are different from those of Hartmann number in the MHD case, there is a strong analogy with the MHD equation (\ref{eqpsi3}). The term on the left-hand side represents the contribution of the Ekman (resp. Hartmann) layer, the first term on the right-hand side corresponds to the effect of topography while the last term is related to shear forces within the two-dimensional bulk flow.  

In terms of boundary-layer structures, equation (\ref{eq-psi-rot}) can be used to estimate the thickness of possible boundary layers arising in rotating flows. For a cavity of constant depth, there can be an equilibrium between the first and last term in (\ref{eq-psi-rot}) on a typical length-scale $E^{1/4}$. This is the thickness of shear boundary layers between regions of different core velocity  [\cite{S57}]. Along western boundaries, the equilibrium between the second and third terms brings a length-scale $E^{1/3}/ \beta$, corresponding to Munk layers [\cite{munk,cw77,w75}]. The equations are not symmetrical between East and West, owing to the term $\beta \partial \psi / \partial s$ in equation (\ref{homogeneous2}), and no boundary layer can develop on eastern boundaries. The development length of a flow along geostrophic contours after a disturbance can be estimated as $E^{-1/2}$ [\cite{w74}].

\section{Circular duct with fringing magnetic field}
\label{fringing}

A long electrically insulating circular duct is submitted to a transverse magnetic field (Fig.~5).
The axis of the pipe is taken to be the $x$-direction, while the direction of the transverse magnetic field is taken to be the $z$ direction. All dimensions are made dimensionless using the diameter of the pipe. From $x=-\infty$ to $x=-2$ the magnetic field intensity is assumed to be uniform. Its value $B_0$ is used as our scale for magnetic field, hence its dimensionless value is unity, while from $x=2$ to $x=\infty$, the dimensionless magnetic field is uniform with an intensity twice as large. Between $x=-2$ and $x=2$, the dimensionless magnetic field is assumed to have the form 
\begin{equation}
B_z = 1.5+\frac{x}{4}+\frac{2}{3 \pi} \sin \left(\frac{\pi x }{ 2} \right) + \frac{1}{12 \pi } \sin ( \pi x ), \label{fieldz}  
\end{equation}
which makes $B_z$ continuous, as well as its first and second derivatives, from $x=-\infty$ to $x=\infty$. According to the straight magnetic field lines approximation, the other components of the magnetic field are assumed to be zero, although this field does not satisfy the condition of a curl-free magnetic field strictly speaking. 

\begin{figure}
\begin{center}
\input{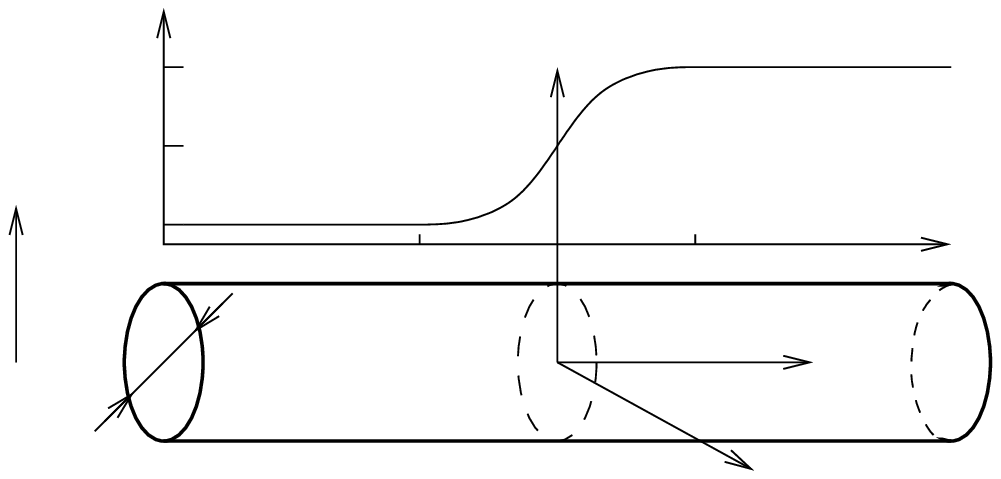}
\caption{Geometry and magnetic field of a circular duct with fringing magnetic field.}
\end{center}
\label{sketch-fring}
\end{figure}

The characteristic surfaces contain the $z$ direction and are best made visible in the ($x$,$y$) plane where they appear as lines. On Fig.~6,
some characteristic surfaces are drawn: they correspond to the case of a circular cylinder and to the distribution defined by the magnetic field (\ref{fieldz}). However, the $x$ and $y$ coordinates are stretched differently. The figure is drawn for the particular case of a duct of length 8, but this is irrelevant since the characteristic surface are independent of $x$ when $ x < -2$ or $  x > 2$. In fact, the length of the cavity used for numerical calculations will be adapted to each value of the Hartmann number (see section \ref{num-fring}) so that it is always longer than the development length of the flow.

\begin{figure}
\begin{center}
\input{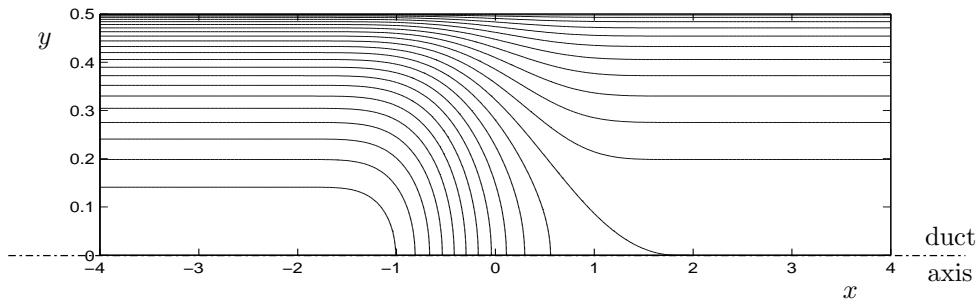}
\caption{Characteristic surfaces, seen on the ($x$,$y$) plane, for half the pipe. The other half $y < 0$ is symmetrical.}
\end{center}
\label{Kar}
\end{figure}

This problem has been initially investigated by Holroyd and Walker [\cite{HetW}] and later by Hua and Walker [\cite{HuaWalker}]. In the next section \ref{num-fring} a two-dimensional numerical calculation of the flow is performed using the two-dimensional model described in section \ref{2dmag}. This is somewhat similar to the work done by Hua and Walker [\cite{HuaWalker}], but higher values of the Hartmann number are investigated here. In section \ref{mod-fring}, an asymptotic model used by Holroyd and Walker is re-expressed in terms of the streamfunctions $\psi$ and $h$ instead of pressure and electrical potential, and is then solved with a higher accuracy (in effect with more terms in a series of Chebyshev polynomials). In this model, it is assumed that the non-uniform region is short enough so that the flow follows the characteristic surfaces exactly on its length-scale. It will be shown that the numerical two-dimensional model and this asymptotic model are in perfect agreement. They both lead to layers of thickness $H\!a^{-1/4}$ and it will be shown that the two-dimensional numerical results approach the asymptotic results when $H\!a^{-1/4}$ is small compared to one. 

\subsection{Numerical two-dimensional results}
\label{num-fring}

The two-dimensional equations (\ref{eqpsi}) and (\ref{eqh}) are solved in this section, except for the two-dimensional viscous term. This is the last term is equation (\ref{eqpsi}) and it has been shown in section \ref{2dmag} that the boundary or free layers arising from it are always as thin as parallel layers, $H\!a^{-1/2}$, hence cannot be modelled correctly by the two-dimensional model. Consequently, this term is removed from the equation and the condition of no-slip on solid boundaries is dropped accordingly. These equations are re-written here, in the form that is solved numerically: 
\begin{eqnarray}
{\bf \nabla}.\left( \frac{B_z}{d^2}  {\cal{F}}. ({\bf \nabla} \psi  \times {\bf e}_z ) \right) {\bf e}_z & = & H\!a \, {\bf \nabla} \left( \frac{B_z}{d} \right) \times {\bf \nabla} h  , \label{eqpsif}\\
{\bf \nabla}. \left( \frac{{\bf \nabla } h }{d} \right) {\bf e}_z 
&=& {\bf \nabla} \left( \frac{B_z}{d} \right) \times {\bf \nabla} \psi. \label{eqhf}
\end{eqnarray}
The symmetry of the problem with respect to the axis $y=0$ is exploited and the domain of integration is only half of the initial domain: $y \geq 0 $, $-L \leq x \leq L$, where $L$ is taken large enough so that the flow is established in the $x$ direction at these two ends. The development length scales as $H\!a^{1/2}$ and this is taken to be the length of the integration domain, $L=0.5 H\!a^{1/2}$. The results show that the flow is independent of $x$ well before the ends. 

\begin{figure}
\begin{center}
\input{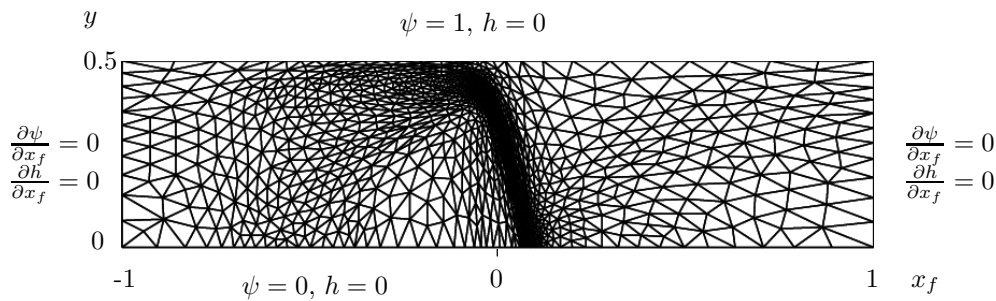}
\caption{Typical mesh, with fewer vertices than used in calculations for clarity, for $H\!a = 10^6$. Boundary conditions for velocity and electric current streamfunctions $\psi$ and $h$  are shown. }
\end{center}
\label{mesh-typ}
\end{figure}

A problem with this domain is that its length is large for large Hartmann numbers and it becomes difficult to build a mesh with a reasonable number of nodes which can cope with the short and large length-scales involved. To overcome that difficulty a change in $x$ coordinate is made, which maps the long physical domain onto a fictitious domain of length 2. The change in $x$-coordinate used here is of the form $x= 2.5 \sinh \left( x_f \sinh ^{-1} (H\!a^{1/2} / 5) \right)$, where $x_f$ is a fictitious $x$ coordinate (see Fig.~7).
This type of stretching creates a distortion that is more and more pronounced towards the ends. 
The following boundary conditions are applied:
on $y=0$, $\psi = 0$ and $h=0$; 
on $y=0.5$, $\psi = 1$ and  $h=0$; 
on $x_f = \pm 1$, $ {\partial \psi}/{\partial x_f} = 0 $ and ${\partial h}/{\partial x} = 0$.

Two types of two-dimensional plots will be shown in this section, none of them involving the artificial $x_f$ coordinate: one type is a general view of a large part of the domain where the $x H\!a^{-1/2}$ will be used while the other type is a limited view of the physical domain, between $x=-2$ and $x=2$, where coordinate $x$ will be used. 

The equations are solved using FreeFem+ [\cite{freefem}], a free-licence software developed at INRIA. This is a two-dimensional finite element solver, coupled to an anisotropic unstructured mesh generator that can automatically refine the mesh where solutions have large gradients. Triangular, first order elements are used. Typically, after a couple of mesh adaptations, the mesh looks like the one displayed on Fig.~7, 
although the meshes used for the calculations contain many more vertices (up to $50,000$).   

\begin{figure}
\hspace*{-0.7 cm} \input{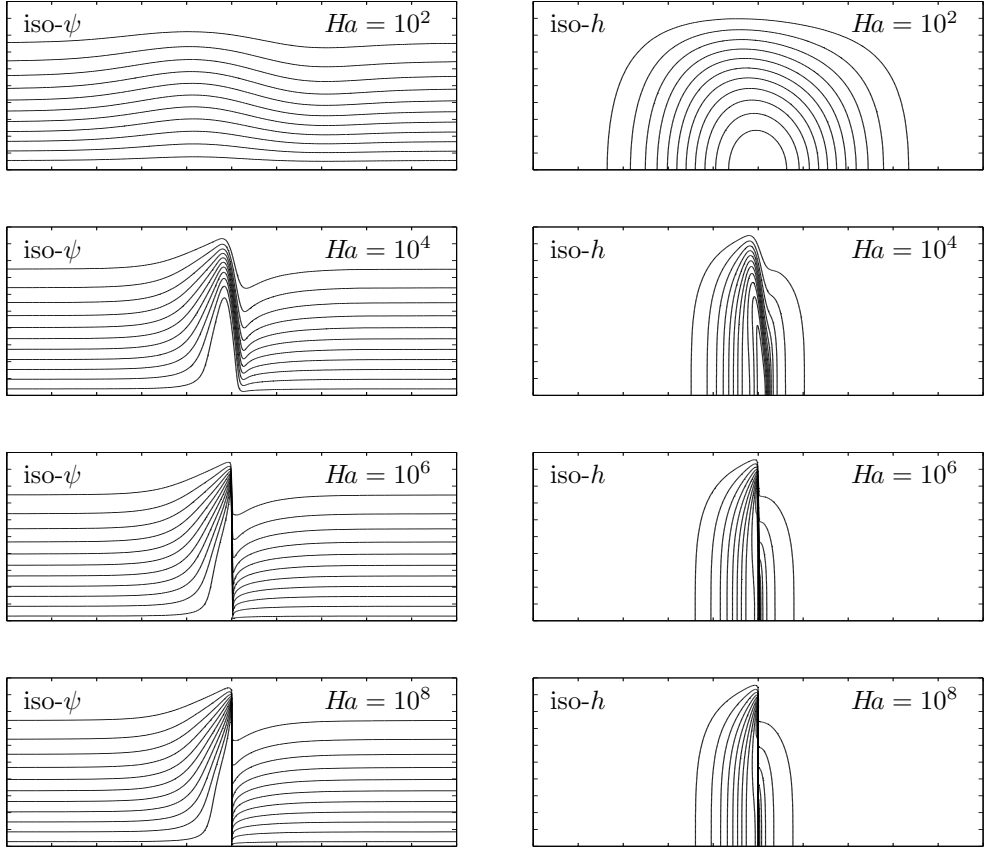}
\begin{center}
\caption{FreeFem calculations: velocity and electric current streamlines, iso-$\psi$ and iso-$h$ on the left and right respectively for 4 values of the Hartmann number, $H\!a = 10^2,\ 10^4, \ 10^6, \ 10^8$. The vertical coordinate is the $y$ coordinate from $y=0$ (centerline of the duct) to $y=0.5$, the horizontal coordinate is $H\!a^{-1/2} x$ from $H\!a^{-1/2} x = - 0.25 $ to $H\!a^{-1/2} x = 0.25 $.} 
\end{center}
\label{freefem}
\end{figure}

It can be seen from Fig.~8
that the large scale structure of the flow and electric current circulation becomes independent of the value of the Hartmann number, at large Hartmann number, {\it i.e.} more than $10^6$ or so. 
Between $H\!a = 10^4$ and $H\!a = 10^6$, some differences in position of the iso-$\psi$ lines may be observed and also of the iso-$h$ lines.  
It is also confirmed from these results that the development length of the flow upstream and downstream, {\it i.e.} along the characteristic surfaces is of order $H\!a^{1/2}$. However, the plots in Fig.~8 do not provide a clear information on the structure of the flow around the region of varying magnetic field. This is due to the fact that the axial coordinate is scaled with a factor $H\!a^{1/2}$ in order to give a global view of the flow, hence the central region, $-2 \leq x \leq 2$ is squeezed enormously in the $x$ direction. The following Fig.~9 
shows the same results as Fig.~8
when only the central part of the duct is singled out. It can also be seen that the flow becomes independent of the Hartmann number, when it is larger than $10^6$. It can also be seen that, on this length-scale of order unity, the velocity and electric current streamlines become closer and closer to the characteristic surfaces (compare with figure \ref{Kar}) at large Hartmann number.  

\begin{figure}
\hspace*{-0.7 cm} \input{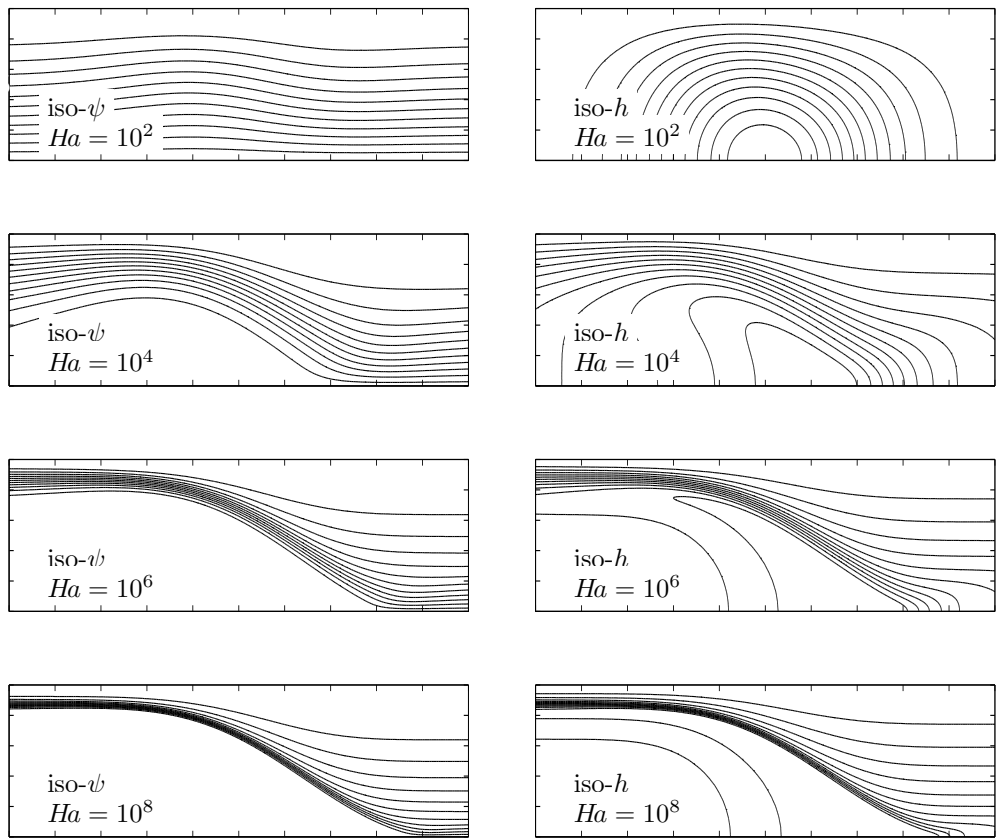}
\begin{center}
\caption{FreeFem calculations: velocity and electric current streamlines, iso-$\psi$ and iso-$h$ on the left and right respectively for 4 values of the Hartmann number, $H\!a = 10^2,\ 10^4, \ 10^6, \ 10^8$. The vertical coordinate is the $y$ coordinate from $y=0$ (centerline of the duct) to $y=0.5$, the horizontal coordinate is $x$ from $x=-2$ to $x=2$.} 
\end{center}
\label{freelocal}
\end{figure}

It is also interesting to look at the velocity profiles of the $x$-component of the velocity field across the duct. On Fig.~10, 
the velocity profile is shown between $y=0$ and $y=0.5$ at the axial positions $x=-2$ and $x=2$ respectively. 

\begin{figure}
\begin{center}
\input{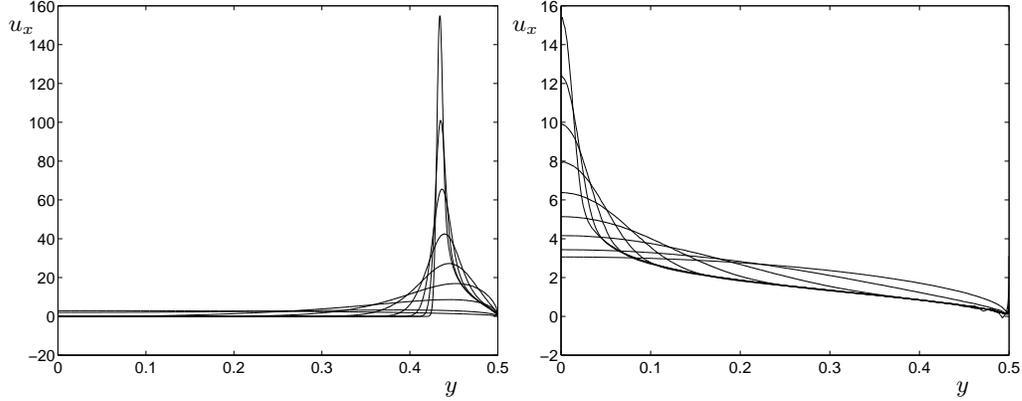}
\caption{FreeFem calculations: velocity profile ($x$ component) at the position $x=-2$ (left) and $x=2$ (right), between $y=0$ (centerline of the duct) and $y=0.5$ (edge of the duct), for  
  $H\!a = 10^2,\ 10^3, \ 10^4, \ 10^5, \ 10^6, \ 10^7, \ 10^8, \ 10^9, \ 10^{10}$.}
\end{center}
\label{plotg}
\end{figure}

One can see that the velocity profile presents a singularity at infinite Hartmann number. This singularity is situated on the characteristic surface which splits in the region of non-uniform magnetic field. Hence, its position is at $y=0$ when the velocity profile is shown at $x=2$, and at $y_c = 0.5 \, \cos (\pi / 6 )$ at $x=-2$. This critical position $y_c$ corresponds to the $y$ position where the depth of the cylinder in the direction of the magnetic field is half the diameter of the duct.  
On Fig.~11, 
the velocity profiles are plotted again, in logarithmic coordinates, to put in evidence the singularity arising at large Hartmann number. On the left ($x=-2$), only the region $y_c < y < 0.5$ is shown and the horizontal coordinate used is $y-y_c$. On the right ($x=2$), the singularity being situated at $y=0$, no change of coordinate is made. It can be seen that, at $x=2$, the velocity profile behaves asymptotically like $y^{-1/2}$ when $H\!a$ increases. On $x=-2$, it looks plausible that the velocity behaves as $(y-y_c)^{-3/4}$ when $H\!a$ increases: it is not as clear as the $-1/2$ exponent at $x=2$, probably because, around $y=y_c$, the depth varies much more rapidly with $y$ than around $y=0$. However, there must be a strong relationship between these exponents on each side of the change of magnetic field if one believes that, on the short length-scale of the change in magnetic field, the streamlines must coincide with the characteristic surfaces
\begin{equation}
\psi [ x = -2 , y_l (K) ] = \psi [x= 2, y_r (K) ], \label{equalpsi}
\end{equation}
where $y_l (K)$ is the $y$ position on the left-hand side of the change of magnetic field for a given value $K$ of the characteristic function $B_z / d $ and where $y_r(K)$ is the corresponding $y$ position on the right ($x=2$). Differentiating expression (\ref{equalpsi}) with respect to $K$, one gets:
\begin{equation}
\frac{d y_l}{d K} \frac{\partial \psi }{\partial y} [ x = -2 , y_l (K) ] = \frac{d y_r}{d K} \frac{\partial \psi }{\partial y} [x= 2, y_r (K) ]. \label{equalvel}
\end{equation}
The derivatives of $y_l$ and $y_r$ with respect to $K$, near the singular characteristic value $K_0 = 2 / 1 = 2$, are the inverse of the derivatives of K with respect to $y$, near $y=y_c$ and $y=0$ respectively. From $K = B_z / d$, with $d=0.5 ( 0.25 - y^2 )^{1/2}$, the result:
\begin{equation}
K = \frac{1}{0.5 ( 0.25 - y_l^2 )^{1/2}} = \frac{2}{0.5 ( 0.25 - y_r^2 )^{1/2}} \label{eqK}
\end{equation}
is obtained, which when differentiated leads to:
\begin{equation}
\frac{d K}{d y_l} = 0.5 y_l (0.25 - y_l^2 )^{-3/2}, \ \ \ \ \ \ \frac{d K}{d y_r} = y_r (0.25 - y_r^2 )^{-3/2}. \label{dKdy}
\end{equation}
This means that, in the neighbourhood of $y_l = y_c$ and $y_r = 0$ respectively:
\begin{equation}
\frac{d K}{d y_l} = 8 \sqrt{3}, \ \ \ \ \ \ \frac{d K}{d y_r} = 8 y_r, \label{dKdy2}
\end{equation}
where for the case of $dK / dy_l$ the only interesting matter is that it has a finite value which is to be considered as uniform near the singularity. 
In contrast, for the case of $dK / dy_r$ a first-order expansion is considered since its value at $y_r = 0$ is zero and cannot be inverted to feed into equation  (\ref{equalvel}). 
Besides, using (\ref{eqK}), it can be shown that near the singularity, the following relationship holds between $y_l$ and $y_r$:
\begin{equation}
y_r^2 \simeq 2 \sqrt{3} (y_l -y _c). \label{eqK2}
\end{equation}
With (\ref{dKdy2}) and (\ref{eqK2}), if it supposed that the the velocity profile in the $x$ direction, $\partial \psi / \partial y_r$, on the right-hand side behaves near $y_r = 0$ as $u_x = a y_r^b$, then equation (\ref{equalvel}) indicates how the velocity should behave on the left-hand side:
\begin{equation}
\frac{\partial \psi}{\partial y_l} (x=-2,y_l) = \sqrt{3} a \left[ 2 \sqrt{3} (y_l -y_c )\right] ^{\frac{b-1}{2}}. \label{relvel}
\end{equation}
This relationship indicate which power law should be found on the left of the singularity given that an exponent $b$ is observed on the right. As the exponent $b=-1/2$ is clearly visible on Fig.~11,
one can infer from (\ref{relvel}) that the singularity in the velocity profile on the left is of the form $(y_l - y_c )^{-3/4}$.

\begin{figure}
\begin{center}
\input{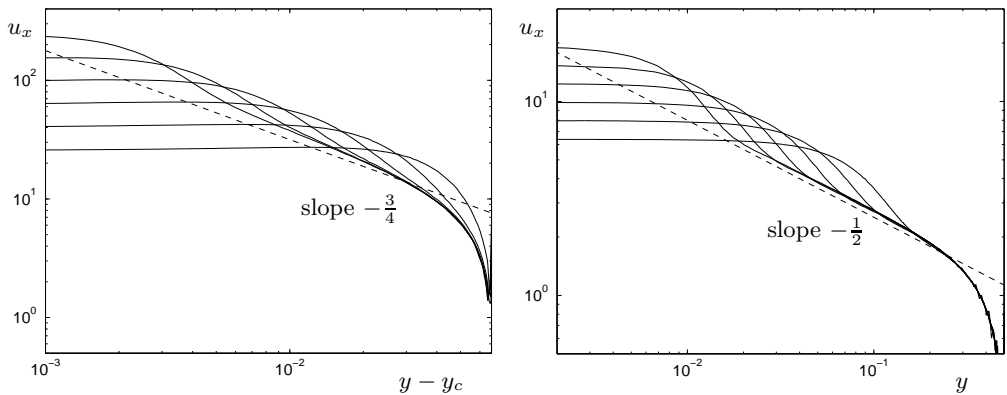}
\caption{FreeFem calculations: velocity profile ($x$ component) at the position $x=-2$ (left), between $y=y_c$ and $y=0.5$ and $x=2$ (right), between $y=0$ and $y=0.5$, for Hartmann numbers $H\!a = 10^6, \ 10^7, \ 10^8, \ 10^9, \ 10^{10}, \ 10^{11}$.} 
\end{center}
\label{plotlog}
\end{figure}

Having established the nature of the singularity developing along this particular characteristic surface which splits as a result of the change of magnetic field, one can also examine how this singularity is smoothed out at high but finite Hartmann number. As predicted in section \ref{2dmag} there is in fact a free shear layer of thickness $H\!a^{-1/4}$. This is because the change in magnetic field occurs on a typical length scale unity, so when the velocity profile is examined at a distance of order unity from the change in magnetic field, a layer of thickness $H\!a^{-1/4}$ is expected. This can be seen on Fig.~10 and 11 
as the singularity is smoothed on a shorter and shorter length scale, $\delta$, as the Hartmann number is increased. One can actually extract quantitatively the `smoothing' thickness out of these velocity profiles. If we assume that the 
velocity $u_x$ is at its maximum $u_{max}$ on a length-scale $\delta$, then the following integrals of the velocity profile can be approximated as: 
\begin{equation}
\int_0^{0.5} u_x^4 dy \sim \delta u_{max}^4, \ \ \ \ \ \int_0^{0.5} u_x^8 dy \sim \delta u_{max}^8, \label{approxdelta}
\end{equation}
as the contribution around the maximum velocity dominates the integrals. 
It then follows from (\ref{approxdelta}) that the following value for $\delta$ is an objective measure of the thickness of the layer which can be computed from the profiles: 
\begin{equation}
\delta = \left[ \frac{\left[ \int_0^{0.5} u_x^4 dy \right] ^{1/4} }{\left[ \int_0^{0.5} u_x^8 dy \right] ^{1/8} } \right] ^8. \label{approxdelta2}
\end{equation}
This value of $\delta$ is plotted on Fig.~12 (left).  
It can be seen that at large Hartmann number, $\delta$ behaves as $H\!a^{-1/4}$. This is a practical application of the general analysis in section \ref{2dmag}. indicating that layers of thickness $H\!a^{-1/4}$ can develop along characteristic surfaces when a change occurs on a length-scale unity. Here the magnetic field changes on a length scale unity and the layer between the stagnant and moving fluids is indeed of thickness $H\!a^{-1/4}$. These findings bring an answer to the discussion started in the conclusion of the paper by Holroyd and Walker [\cite{HetW}]. In their conclusion, about the `{\it nature of the velocity profile near the moving fluid/stagnant fluid boundary in the non-uniform region}', the two authors express different views: `{\it J.S.W believes the velocity gradients to be large yet ${\cal{O}}(1)$ while R.J.H. believes that a shear layer of thickness ${\cal{O}}(H\!a^{-1/2})$ might separate the moving and stagnant fluid}'.  
The answer proposed here is intermediate, with a layer of thickness $H\!a^{-1/4}$. In addition, it is observed that the asymptotic velocity profile itself diverges near $y_c$, not only its gradient.

\begin{figure}
\begin{center}
\input{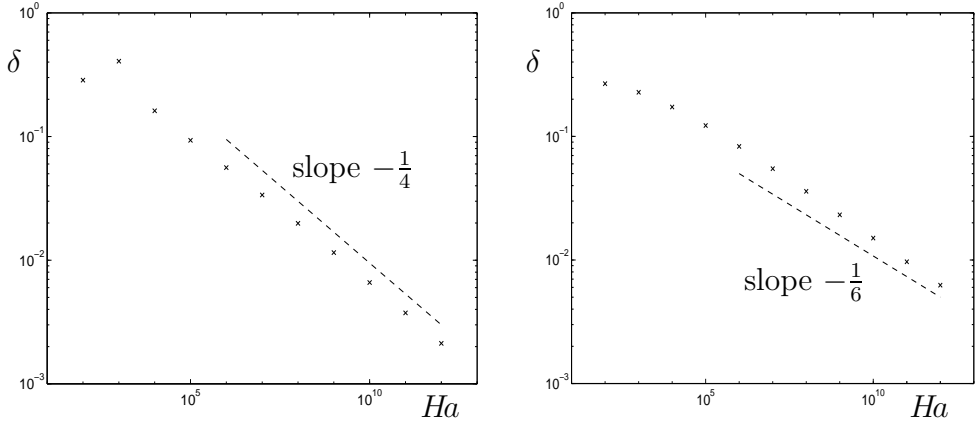}
\caption{Expression (\ref{approxdelta2}) is plotted for Hartmann number from $10^2$ to $10^{12}$, based on the velocity profile at $x=-2$ (left) and $x=2$ (right). The dashed lines are straight lines of slope $-1/4$ and $-1/6$ for comparison.}
\end{center}
\label{figdelta}
\end{figure}

If one observes now the velocity profile on the right-hand side of the change in magnetic field, one can also measure the size of the free-shear layer at $y=0$, lying on the dividing characteristic surface. Using the same expression (\ref{approxdelta2}), the thickness of this layer is plotted on Fig.~12 (right).
It can be seen that, at high Hartmann number, the thickness decreases as $H\!a^{-1/6}$. This new exponent arises because the characteristic function $K$ shows an extremum at $y=0$. Indeed, the power law $H\!a^{-1/4}$ was derived under the assumption that $K$ has a finite gradient $G$. Here, $G$ is not constant at the scale of the layer, but is a linear function of $y$, $G=c z$, where $c$ is a constant. When substituting this expression for $G$ in equation (\ref{eqpsi5}), it is possible to carry on a similar scaling analysis as in \ref{2dmag} to find out that a shear layer of thickness $H\!a^{-1/6}$ should develop near $y=0$ for the axial position $x=2$.

\subsection{The model of Holroyd and Walker revisited} 
\label{mod-fring}
 
The method of eigenfunction expansion used by Holroyd and Walker [\cite{HetW}] (directly adapted itself from a paper by Walker and Ludford [\cite{WalkerLudford}]) has been followed with two changes. First, the problem is solved here using the streamfunctions $\psi$ and $h$, rather the the pressure and electric potential $p$ and $\phi$, and secondly, as many as 1000 eigenvectors have been used while Holroyd and Walker could only compute 60 at the time of their publication. This increase in computer power is necessary to observe the results found in the previous section with FreeFem. In this method, the magnetic field is modelled as a pure step function and the velocity profiles are calculated at a distance $x=2$ upstream and downstream of the discontinuity, and plotted in linear (Fig. 13) and logarithmic (Fig.~14) coordinates.  

\begin{figure}
\begin{center}
\input{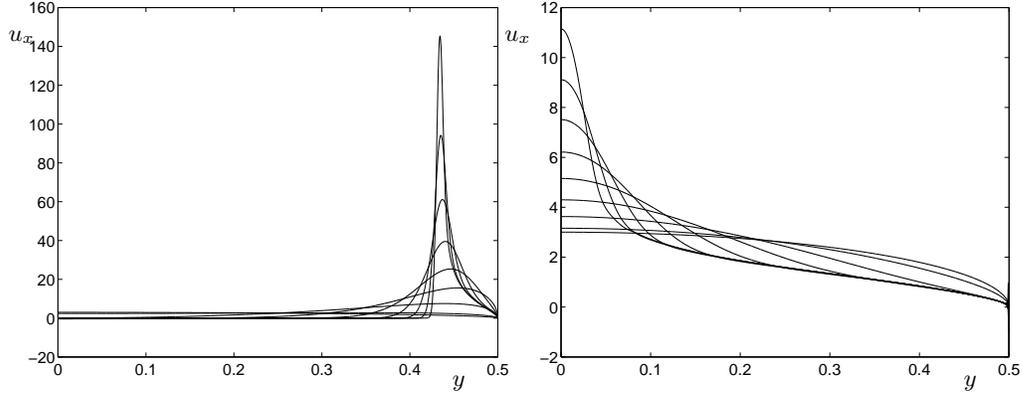}
\caption{Eigenfunction expansions: velocity profile ($x$ component) at the position $x=-2$ (left) and $x=2$ (right), between $y=0$ (centerline of the duct) and $y=0.5$ (edge of the duct), for  
  $H\!a = 10^2,\ 10^3, \ 10^4, \ 10^5, \ 10^6, \ 10^7, \ 10^8, \ 10^9, \ 10^{10}$.}
\end{center}
\label{plotgcheb}
\end{figure}

\begin{figure}
\begin{center}
\input{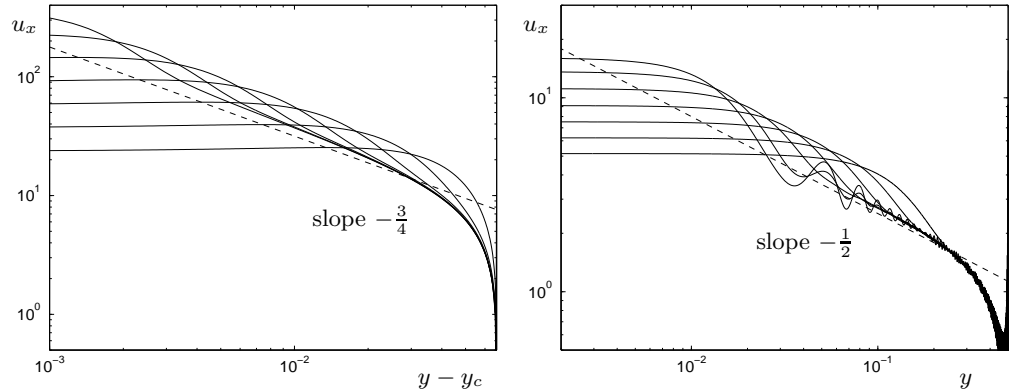}
\caption{Eigenfunction expansions: velocity profile ($x$ component) at the position $x=-2$ (left), between $y=y_c$ and $y=0.5$ and $x=2$ (right), between $y=0$  and $y=0.5$, for Hartmann numbers $H\!a = 10^6, \ 10^7, \ 10^8, \ 10^9, \ 10^{10}, \ 10^{11}, \ 10^{12}$. On the left, the abscissa is the $y$ coordinate measured from the singularity $y-y_c$.} 
\end{center}
\label{plotlogcheb}
\end{figure}

The results of this modelling are very similar to those obtained by finite element analysis (Fig.~10 and 11). This can be seen as a confirmation that the approach of Holroyd and Walker was correct. 
The results are not identical though for two reasons. First, at $H\!a=10^{12}$ and to a lesser extend at $H\!a=10^{11}$, one can see numerical oscillations on the right of Fig.~14, showing that 1000 eigenmodes is not sufficient. Secondly, the peak values in shear layers are not identical because $x=-2$ or $x=2$ does not have the same significance in both models. They are the ends of the gradient region of magnetic field in the finite element modelling while they are exact positions with respect to a sharp change in magnetic field intensity in the method of matched expansions.

\section{The entry problem in rectangular ducts}
\label{entry}

The so-called entry problem is another typically relevant issue regarding the two-dimensional MHD flow structure. This phenomenon concerns the changes in a duct flow when the intensity of an imposed transverse magnetic field varies along the direction of the flow, for instance when a duct flow enters (or leaves) a magnet, hence the term `entry problem'. In electrically insulating rectangular ducts, it is observed that the flow remains quasi two-dimensional but concentrates near the two parallel walls to the magnetic field while becoming very weak in the middle of the duct: because of the no-slip condition at the wall, this has been described as an M-shape velocity profile when plotted from one parallel wall to the other. An entirely equivalent problem is that of a flow in a uniform magnetic field but with a variable cross-section area along the duct. Indeed, only changes in the value of the characteristic function $B_z / d$ are relevant. Here, we choose to consider the case of a uniform depth $d=1$ and a variable magnetic field intensity $B_z$ as a function of $x$, the longitudinal coordinate along the duct axis. As before $z$ is the coordinate in the direction of the magnetic field and $y$ is the other coordinate in the plane of the cross-section.

This problem will be treated using equations (\ref{eqpsi}) and  (\ref{eqh}), and our choice leads to simplified governing equations:
\begin{eqnarray}
H\!a \, \left[ {\bf \nabla}.\left( {B_z} {\bf \nabla} \psi \right) \right] \  {\bf e}_z & = & H\!a^2 \, {\bf \nabla} {B_z} \times {\bf \nabla} h  + \left[ \left( {\bf \nabla} ^2 \right) ^2 \psi \right] {\bf e}_z, \label{eqpsis}\\
\left[ {\bf \nabla} ^2  h \right] \ {\bf e}_z 
&=& {\bf \nabla} {B_z} \times {\bf \nabla} \psi. \label{eqhs}
\end{eqnarray}
Then, following our analysis in section \ref{}, the last term of equation (\ref{eqpsis}) -- of viscous origin -- will be neglected as it leads to thin parallel layers, too thin for the condition of two-dimensional flow to apply. Assuming that its role will be simply to ensure the no-slip condition, this term is removed while the no-slip condition is dropped.  
The set of equations becomes:
\begin{eqnarray}
\left[ {\bf \nabla}.\left( {B_z} {\bf \nabla} \psi \right) \right] \  {\bf e}_z & = & H\!a \, {\bf \nabla} {B_z} \times {\bf \nabla} h  , \label{eqpsis2}\\
\left[ {\bf \nabla} ^2  h \right] \ {\bf e}_z 
&=& {\bf \nabla} {B_z} \times {\bf \nabla} \psi. \label{eqhs2}
\end{eqnarray}
These equations are solved in section \ref{num-rect}, for the case of a duct flow with a ramp of transverse magnetic field. Before, in section \ref{local-rect}, a local analytical solution for equations (\ref{eqpsis2}) and (\ref{eqhs2}) is derived corresponding to boundary layers analogous to Stommel layers for rotating flows. 
Finally, in section \ref{section-rect}, an attempt is made to calculate the flow in the duct cross-section in a region of axially varying $B_z$: it is compared to and gives support to the two-dimensional results. 

\subsection{Analytical local solution}
\label{local-rect}

One considers here a region near a side wall, where the gradient of magnetic field $B_z$ is denoted $G$. In the Laplacian terms, the normal derivative along $y$ is assumed to be dominant compared to the axial $x$ derivative. Equations (\ref{eqpsis2}) and (\ref{eqhs2}) can be written:
\begin{eqnarray}
 {B_z} \frac{ \partial ^2 \psi }{\partial y^2} & = & H\!a \, G \, \frac{\partial h}{\partial y} \label{eqpsis3}\\
\frac{ \partial ^2 h }{\partial y^2} & = &  
G\, \frac{\partial \psi}{\partial y}. \label{eqhs3}
\end{eqnarray}
Since $B_z$ and $G$ are functions of $x$ only, these equations are ordinary linear differential equations in $y$. They admit elementary solutions of the form:
\begin{equation}
\exp \left( y \sqrt{ \frac{H\!a G^2}{2 B_z} }   \right), \ \ \ \ \exp \left( -y \sqrt{  \frac{H\!a G^2}{2 B_z} }   \right), \label{expon}
\end{equation}
explaining that the flow is confined near the lateral walls in a boundary layer of typical thickness $H\!a^{-1/2}G^{-1}$, when the scale for magnetic field strength is chosen locally ($B_z =1$). As long as the gradient of magnetic field $G$ is very small compared to one, the two-dimensional structure of the layer is guaranteed and its velocity and electric current density must be exponential.  

\subsection{Numerical two-dimensional results}
\label{num-rect}

A duct of uniform dimensionless depth unity, width equal to 2, and length 8 is considered. The upstream and downstream region occupy each one fourth of the length of the duct and have a uniform transverse magnetic field, of intensity $1-2G$ and $1+2G$ respectively, while the middle part is submitted to a uniform gradient $G$, $B_z = 1 + G x$. On the lateral walls, $y=\pm 1$, the electrically insulating condition can be written $h=0$ and the condition of impermeable walls is written $\psi =0$ on the lower wall at $y=-1$ and $\psi = 1$ at $y=1$\footnote{The condition $\psi=1$ can be changed for an arbitrary constant without affecting the nature of the solution as we are treating a linear problem.}. At the ends of our duct, we apply the condition that $h$ and $\psi$ have no normal gradient. This is consistent with the fact the flow is nearly fully established at the ends.

The streamlines of the computed solution are shown on Fig.~15 
for $H\!a = 10^5$ and three values of the gradient of magnetic field applied, $G=0.05$, $G=0.1$ and $G=0.2$. It can be seen that the side layers become thicker as $G$ decreases. In the real three-dimensional case, it is expected that a parallel layer will exist, as a sublayer within this thicker side layer. 
The exponential velocity distribution (\ref{expon}) has been compared with the calculated distribution with FreeFem: the agreement is excellent.
\begin{figure}
\begin{center}
\input{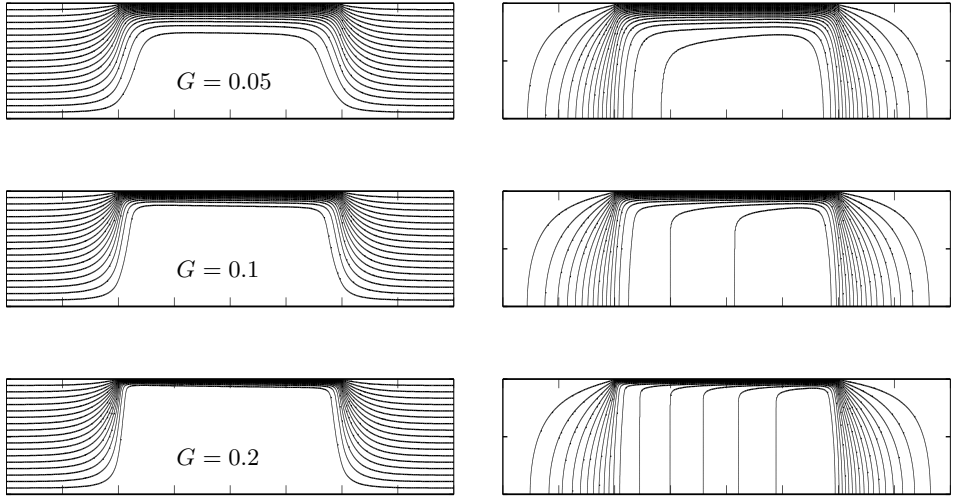}
\caption{Velocity streamlines (left) and electric current streamlines (right) resulting from the two-dimensional numerical analysis for $H\!a = 10^5$ and three values of $G$, $G = 0.05$ on the top, $G=0.1$ in the middle and $G=0.2$ at the bottom. The vertical coordinate is the $y$ coordinate from $y=0$ (centerline of the duct) to $y=1$, the horizontal coordinate is $x$ from $x=-4$ to to $x=4$. The symmetrical part of the cavity ($-1 < y < 0$) is not represented here.}
\end{center}
\label{str-rect}
\end{figure}

\subsection{Local solution in a cross-section}
\label{section-rect}

As we have seen in section \ref{local-rect}, the solution of the duct flow with a gradient of magnetic field is local: it is sufficient to know the local value of the magnetic field $B_z$ and the local value of the gradient $G$ to determine completely the boundary layer solution, providing $B_z$ and $G$ do not vary significantly on an axial length-scale comparable to the layer thickness. As a consequence, it is expected that the three-dimensional flow can be computed locally in a cross section, within some simplifying assumptions.

Let us assume that the flow will be mainly streamwise, and that its streamwise gradient is small compared to its gradient in the cross-section. Navier-stokes equation can be written in the streamwise direction: 
\begin{equation}
-\frac{\partial p}{\partial x} + H\!a^2 j_y B_z + \nabla_S ^2 u_x = 0, \label{navier-streamwise}
\end{equation}
where $\nabla_S ^2$ is the Laplacian operator in the cross-section. This equation (\ref{navier-streamwise}) will be one of our three equations within the cross-section, the $u_x$ equation, and we shall now derive equations for $\phi$ and ${\partial p}/{\partial x}$. The unknown $j_y$ in (\ref{navier-streamwise}) can be expressed as a function of $\phi$ and $u_x$ according to Ohm's law: $j_y = -\partial \phi / \partial y - u_x B_z$.

The equation for $\phi$ is obtained as usual by taking the divergence of Ohm's law:
\begin{equation}
\nabla ^2 \phi = {\bf B}.  {\bf  \nabla } \times {\bf u}, \label{div_ohm}
\end{equation}
With a dominant velocity $x$-component, and with gradients predominantly in the cross-section, this equation can be written:
\begin{equation}
\nabla_S ^2 \phi = - B_z \frac{\partial u_x}{\partial y}, \label{div_ohm_S}
\end{equation}

An equation for ${\partial p}/{\partial x}$  is a less standard object. It is obtained through a few steps. First, we consider the curl of the Navier-Stokes equation:
\begin{equation}
H\!a^2 B_z \frac{\partial {\bf j} }{\partial z} - H\!a^2 G j_x {\bf e}_z + \nabla ^2 ({\bf \nabla } \times {\bf u}) = 0, \label{curl-navier}
\end{equation}
As a next step, we shall write the x-component of the curl of equation (\ref{curl-navier}): 
\begin{equation}
H\!a^2 B_z \frac{\partial ({\bf \nabla} \times {\bf j})_x }{\partial z} - H\!a^2 G \frac{\partial j_x}{\partial y} - ( \nabla ^2 ) ^2 {\bf u}_x = 0, \label{curl-curl-navier}
\end{equation}
To make progress, we need to derive the curl of Ohm's law:
\begin{equation}
{\bf \nabla} \times {\bf j} = B_z \frac{\partial {\bf u}}{\partial z} - u_x G {\bf e}_z, \label{ohm-X}
\end{equation}
where ${\bf e}_z$ is the unit vector in the $z$ direction. 
Substituting into (\ref{curl-curl-navier}), and assuming again that gradients are essentially due to variations with the cross-section, we get:
\begin{equation}
H\!a^2 B_z^2 \frac{\partial ^2 u_x }{\partial z^2} - H\!a^2 G^2 u_x - ( \nabla _S ^2 ) ^2 u_x = 0, \label{curl-curl-navier2}
\end{equation}
On the other hand, one can derive a similar equation by taking directly the cross-section Laplacian of equation (\ref{navier-streamwise}):
\begin{equation}
- \nabla _S ^2 \frac{\partial p}{\partial x} +  H\!a^2 B_z^2 \nabla _S ^2 j_y  - ( \nabla _S ^2 ) ^2 u_x = 0. \label{curl-curl-navier3}
\end{equation}
Taking the curl of equation (\ref{ohm-X}) and considering its $y$ component leads to:
\begin{equation}
- \nabla_S ^2 j_y = B_z \frac{\partial ^2 u_x}{\partial z^2}, \label{curl-ohm-X}
\end{equation}
as variations along $x$ are neglected compared to variations within the cross-section.  
Substituting into (\ref{curl-curl-navier3}) leads to:
\begin{equation}
- \nabla _S ^2 \frac{\partial p}{\partial x} -  H\!a^2 B_z^2 \frac{\partial ^2 u_x }{\partial z^2}  - ( \nabla _S ^2 ) ^2 u_x = 0. \label{curl-curl-navier4}
\end{equation}
The difference between equation (\ref{curl-curl-navier2}) and (\ref{curl-curl-navier4}) gives an equation for $\partial p / \partial x$:
\begin{equation}
- \nabla _S ^2 \frac{\partial p}{\partial x} =  H\!a^2 G^2 u_x . \label{eqp}
\end{equation}

In summary, we have a system of three equations for three unknowns:
\begin{eqnarray}
\nabla_S ^2 u_x &=& \frac{\partial p}{\partial x} + H\!a^2 B_z \frac{ \partial \phi }{\partial y} +  H\!a^2  u_x B_z^2 , \label{navier-streamwise-f} \\
\nabla_S ^2 \phi &=& - B_z \frac{\partial u_x}{\partial y}, \label{div_ohm_S-f} \\
\nabla _S ^2 \frac{\partial p}{\partial x} &=&  - H\!a^2 G^2 u_x . \label{eqp-f}
\end{eqnarray}

This set of equations is solved for the same geometry as considered in section (\ref{num-rect}). The cross-section has dimension 1 along the magnetic field direction and 2 in the transverse direction. Because of the symmetries of the configuration, we solve this problem in a quarter of the cross-section. 

\begin{figure}
\begin{center}
\input{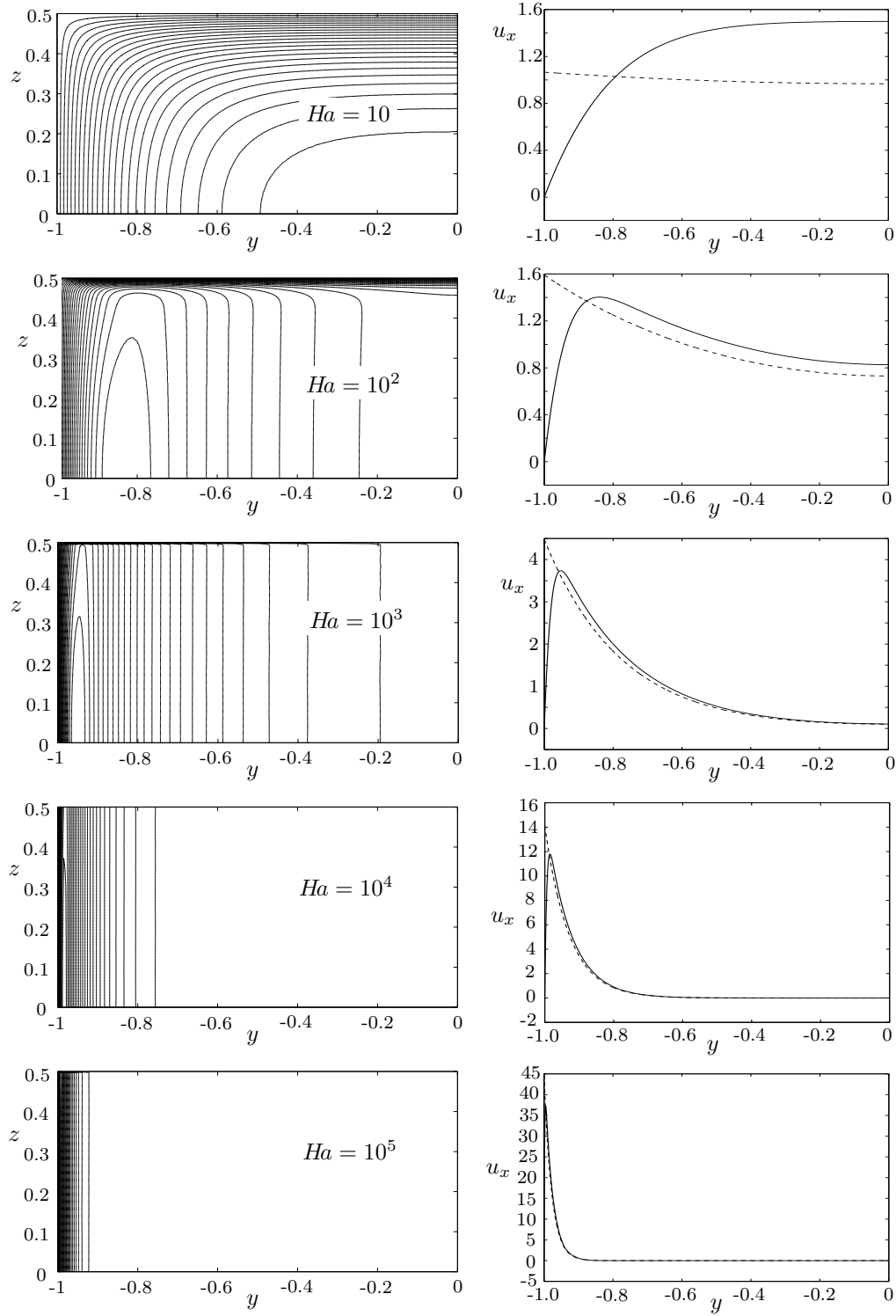}
\caption{Numerical calculation of the flow in the cross-section of a rectangular duct. Using symmetries, only a quarter of the cross-section is analyzed. Isovalues of the streamwise velocity (left) for $H\!a = 10$, $10^2$, $10^3$, $10^4$ and $10^5$, with $G = 0.2$. Velocity profile at $z=0$ (right) for the same parameters, compared to the analytical solution (\ref{cosh}).}
\end{center}
\label{sect-rect}
\end{figure}

Finally, we need to specify boundary conditions for $u_x$, $\phi$ and ${\partial p}/{\partial x}$. The conditions on $u_x$ and $\phi$ are obvious, with $u_x$ vanishing on the walls, and $\partial \phi / \partial n = 0$ as electrically insulating boundaries are considered. Only approximate boundary conditions can be found for ${\partial p}/{\partial x}$; the assumption that this variable is going to be nearly independent of $z$ is made. This assumption is true within the core, it is also quite true within Hartmann layers as they are so thin that pressure cannot change significantly across them, and we also assume that three-dimensional parallel layers ($H\!a^{-1/2}$) are thin enough so that they cannot cause large $z$-dependence of $\partial p / \partial x$. Then, integrating equation (\ref{eqp-f}) across the whole cross-section tells us that, along parallel walls, the normal derivative of ${\partial p}/{\partial x}$ is proportional to the total volume flow rate though the duct, Q:
\begin{equation}
\frac{ \partial}{\partial y} \left( \frac{\partial p}{\partial x} \right) = \pm H\!a^2 G^2 \frac{Q}{2}, \label{bc} 
\end{equation}
while this derivative vanishes on Hartmann layer walls. An arbitrary value $Q=2$ will be used in the calculations so that the mean velocity is unity.

The distribution of axial velocity, electric potential and axial pressure gradient was calculated with FreeFem++, an improved version of FreeFem+ where it is possible to used triangular second order finite elements. This proved necessary as Hartmann layers were resolved for Hartmann number up to $10^5$. In this cross-section modelling, Hartmann layers have to be very well resolved as the electric current flowing through them is a key ingredient in the establishment of $H\!a^{-1/2}G^{-1}$ layers.  

Some results are shown in Fig.~17,
for a fixed gradient $G=0.2$, and increasing Hartmann numbers, from $10$ to $10^5$. As expected, shear layers of thickness $H\!a^{-1/2}G^{-1}$ develop and parallel layers of thickness $H\!a^{-1/2}$ are sublayers: their role is indeed to bring the velocity to zero at the wall.   
The velocity profile at $z=0$ is compared with the analytical exponential solution. By symmetry, using expressions (\ref{expon}) and for a total volume flow rate equal to 2 ($Q=2$), the analytical expression (dashed line in Fig.~17) is:
\begin{equation}
u_x = \sqrt{ \frac{H\!a G^2}{2 B} } \frac{ {\rm cosh} \left( \sqrt{ \frac{H\!a G^2}{2 B} } y  \right) }{{\rm sinh} \left( \sqrt{ \frac{H\!a G^2}{2 B} } \right) }. \label{cosh}
\end{equation}
The agreement is very good, the small departure being due to the flow deficit in parallel layers at large Hartmann number and to the deficit in Hartmann layers at smaller Hartmann numbers. This shows that our local problem in a cross-section (equations (\ref{navier-streamwise-f}), (\ref{div_ohm_S-f}) and (\ref{eqp-f})) is a good representation of MHD flows for entry problems. It is also a direct confirmation that the results of two-dimensional modelling can be confirmed from a model where Hartmann layers are effectively calculated. 

\section{Concluding remarks}

The structure of MHD flows under large Hartmann numbers has been investigated for an arbitrary geometry of the container and magnetic field distribution. 
A set of two equations governing the two-dimensional mass flux streamfunction $\psi$ and electric charge flux streamfunction $h$ has been used, however it would be equivalent to consider other variables such as pressure $p$ and electric potential $\phi$ as in [\cite{HetW}]. The reason why mass and electric fluxes are preferred is that this approach shows clearly which properties of Hartmann and Ekman layer play a key role in the two-dimensional equations (see section \ref{2d}). In the case of rotating flows, the key property of Ekman layers is to carry a mass flux in the direction perpendicular to the core velocity. In the case of MHD flows, the key property of Hartmann layers is to carry an electric current in the direction perpendicular to the core velocity.

The originality of the present work consists largely in the scaling analysis of these MHD two-dimensional equations. It has been shown that shear layers of thickness $H\!a^{-1/4}$ can develop. In addition, the analogy between rotating flows and MHD flows has been further developed, so that for instance it is now clear that the analogue of Stommel layers are those analyzed by Walker and Ludford [\cite{WL72}] of thickness of order $H\!a^{-1/2} G^{-1}$. 

It may be that some of the results presented in this paper, especially those concerning layers of thickness $H\!a^{-1/4}$, will never realistically apply. In industrial MHD situations, one may envisage Hartmann numbers up to $10^5$, with 10 Tesla and 10$\, $cm length-scale for a liquid metal. However, even in this case, it may be interesting to know what the flow structure would be at higher Hartmann numbers, as a tendency towards this structure will already exist at lower Hartmann numbers. In the astrophysical and geophysical context, it is easy to find higher estimates of Hartmann numbers. For instance, in the liquid Earth core, one can reach $H\!a \sim 10^7$ or $H\!a \sim 10^8$ depending on whether this is based on a poloidal ($B \simeq 5\, 10^{-4}$) or a toroidal ($B \simeq 5\, 10^{-3}$) estimate for the magnetic field intensity (see [\cite{cbjnm02}]). Nevertheless, rotating effects, inertial effects and transport of magnetic field are so important that the Hartmann number is not the key dimensionless parameter. 

A number of possible extensions of this work can be envisaged. First, in this paper, the use of the straight magnetic lines approximation has been made and a more accurate two-dimensional model could be obtained by taking into account of the true magnetic field distribution. Although the results would be changed quantitatively, it is not expected that the flow structure would be altered significantly. 
Secondly, the present analysis can be extended to analyse free-surface flows. The position of the free surface then becomes an additional unknown and the nature of Hartmann layers changes : they become much less active in the sense that the electric current flowing along them is drastically reduced compared to a wall-bounded Hartmann layer. Shear layers of a new type are expected as the balance in our equation (\ref{eqpsi3}) will be changed.
A third possible extension would be to consider the combined effect of rotation and magnetic field. This is typical of astrophysical or planetary dynamics. When rotation and magnetic field have an identical direction, there is certainly a predominantly two-dimensional flow. In the general case of different orientations, it is far from obvious that such a two-dimensional dynamics will develop, unless one effect (rotation or magnetic field) is dominant compared to the other one.  

The analogy with rotating flows has been used as an example to follow in the MHD case. However, if one takes the opposite view and wonder what should be the analogous of $H\!a^{-1/4}$ layers, one can go back to equation (\ref{homogeneous2}) for the homogeneous model. If one considers a change on a length-scale unity 
along a geostrophic contour, one may wonder what the thickness $\delta$ of the shear layer along that contour should be. In Oceans, as $\beta $ is small, the answer is that it should be of order $E^{1/4}\beta ^{-1/2}$ and that it is the result of the competition between topographic effects and Ekman pumping effects. This layer is thicker than the $E^{1/4}$ Stewartson layer, which corresponds to the balance between bulk viscous effects and Ekman pumping. Are there shear layers, lying along geostrophic contours, of thickness of order $E^{1/4}\beta ^{-1/2}$? 

\begin{acknowledgements}
{{\bf Acknowledgments: }This work was started at the Department of Engineering in Cambridge and the author is grateful to his colleague Martin Cowley for numerous discussions on this and related topics and for helpful comments on the paper. Thanks are due also to Andrew Soward, University of Exeter, for bringing Munk and Stommel boundary layers to the author's attention.}
\end{acknowledgements}

\bibliography{bib}

\end{document}